\documentclass[iop]{emulateapj}
\usepackage{apjfonts,rotating}

\def\asca{{\itshape ASCA\/}}

\def\chandra{{\itshape Chandra\/}}

\def\hst{{\itshape HST\/}}

\def\spitzer{{\itshape Spitzer\/}}

\def\xray{\hbox{X-ray}}

\def\etal{{et\,al.}}
\def\ae{{\ttfamily AE}}

\def\ltsima{$\; \buildrel < \over \sim \;$}
\def\simlt{\lower.5ex\hbox{\ltsima}}
\def\gtsima{$\; \buildrel > \over \sim \;$}
\def\simgt{\lower.5ex\hbox{\gtsima}}
\def\kms{\ifmmode{~{\rm km~s^{-1}}}\else{~km s$^{-1}$}\fi}
\def\lsim{\lower0.3em\hbox{$\,\buildrel <\over\sim\,$}}
\def\gsim{\lower0.3em\hbox{$\,\buildrel >\over\sim\,$}}
\def\lbsol{$L_{B,\odot}$}
\def\lksol{$L_{K,\odot}$}

\def\h2{H$_2$}
\def\flux{erg~cm$^{-2}$~s$^{-1}$}
\def\lum{erg~s$^{-1}$}

\def\arcsec{\mbox{$^{\prime\prime}$}}
\def\arcmin{\mbox{$^\prime$}}

\def\aap{A\&A}
\def\apj{ApJ}
\def\apjl{ApJL}
\def\apjs{ApJS}
\def\aj{AJ}
\def\mnras{MNRAS}
\def\araa{ARA\&A}

\begin{document}

\shortauthors{LEHMER ET AL.}
\shorttitle{Dependence of Field LMXB XLFs with Stellar Age}

%
\title{The X-ray Luminosity Functions of Field Low Mass X-ray Binaries in Early-Type Galaxies: Evidence for A Stellar Age Dependence}
%

\author{
B.~D.~Lehmer,\altaffilmark{1,2}
M.~Berkeley,\altaffilmark{2,3}
A.~Zezas,\altaffilmark{4,5,6}
D.~M.~Alexander,\altaffilmark{7}
A.~Basu-Zych,\altaffilmark{2,8}
F.~E.~Bauer,\altaffilmark{9,10}
W.~N.~Brandt,\altaffilmark{11,12}
T.~Fragos,\altaffilmark{5}
A.~E.~Hornschemeier,\altaffilmark{2} 
V.~Kalogera,\altaffilmark{13} 
A.~Ptak,\altaffilmark{2} 
G.~R.~Sivakoff,\altaffilmark{14}
P.~Tzanavaris,\altaffilmark{1,2}
\& M.~Yukita\altaffilmark{1,2}
}

\altaffiltext{1}{The Johns Hopkins University, Homewood Campus, Baltimore, MD
21218, USA}
\altaffiltext{2}{NASA Goddard Space Flight Center, Code 662, Greenbelt, MD
20771, USA} 
\altaffiltext{3}{Institute for Astrophysics and Computational Sciences,
Department of Physics, The Catholic University of America, Washington, DC
20064, USA}
\altaffiltext{4}{Physics Department, University of Crete, Heraklion, Greece}
\altaffiltext{5}{IESL, Foundation for Research and Technology, 71110 Heraklion,
Crete, Greece}
\altaffiltext{6}{Harvard-Smithsonian Center for Astrophysics, 60 Garden Street,
Cambridge, MA 02138, USA}
\altaffiltext{7}{Department of Physics, University of Durham, South Road,
Durham DH1 3LE, UK}
\altaffiltext{8}{Center for Space Science and Technology, University of
Maryland Baltimore County, 1000 Hilltop Circle, Baltimore, MD 21250, USA}
\altaffiltext{9}{Pontificia Universidad Catolica de Chile, Departamento de
Astronomia y Astrofisica, Casilla 306, Santiago 22, Chile}
\altaffiltext{10}{Space Science Institute, 4750 Walnut Street, Suite 205,
Boulder, Colorado 80301, USA}
\altaffiltext{11}{Department of Astronomy and Astrophysics, Pennsylvania State
University, University Park, PA 16802, USA}
\altaffiltext{12}{Institute for Gravitation and the Cosmos, Pennsylvania State
University, University Park, PA 16802, USA}
\altaffiltext{13}{Department of Physics and Astronomy, Northwestern University,
2145 Sheridan Road, Evanston, IL 60208, USA}
\altaffiltext{14}{Department of Physics, University of Alberta, CCIS 4-183
Edmonton, AB T6G 2E1, Canada}
%

%
\begin{abstract}
%

We present direct constraints on how the formation of low-mass \xray\ binary
(LMXB) populations in galactic fields depends on stellar age.  In this pilot
study, we utilize \chandra\ and {\it Hubble Space Telescope} (\hst) data to
detect and characterize the \xray\ point source populations of three nearby
early-type galaxies: NGC~3115, 3379, and 3384.  The luminosity-weighted stellar
ages of our sample span \hbox{$\approx$3--10~Gyr}.  X-ray binary population
synthesis models predict that the field LMXBs associated with younger stellar
populations should be more numerous and luminous per unit stellar mass than
older populations due to the evolution of LMXB donor star masses.  Crucially,
the combination of deep \chandra\ and \hst\ observations allows us to test
directly this prediction by identifying and removing counterparts to \xray\
point sources that are unrelated to the field LMXB populations, including LMXBs
that are formed dynamically in globular clusters, Galactic stars, and
background AGN/galaxies.  We find that the ``young'' early-type galaxy NGC~3384
(\hbox{$\approx$2--5~Gyr}) has an excess of luminous field LMXBs ($L_{\rm X}
\simgt$~\hbox{(5--10)}~$\times 10^{37}$~\lum) per unit $K$-band luminosity
($L_K$; a proxy for stellar mass) than the ``old'' early-type galaxies NGC~3115
and 3379 (\hbox{$\approx$8--10~Gyr}), which results in a factor of
$\approx$\hbox{2--3} excess of $L_{\rm X}/L_K$ for NGC~3384.  This result is
consistent with the \xray\ binary population synthesis model predictions;
however, our small galaxy sample size does not allow us to draw definitive
conclusions on the evolution field LMXBs in general.  We discuss how future
surveys of larger galaxy samples that combine deep \chandra\ and \hst\ data
could provide a powerful new benchmark for calibrating \xray\ binary population
synthesis models.

%
\end{abstract}
%


%
\section{Introduction}
%

The {\itshape \chandra\ \xray\ Observatory} (\chandra) has contributed greatly to our
understanding of compact objects (e.g., neutron stars and black holes) through
the study of \xray\ binary populations in normal galaxies.  Through correlative
studies involving \xray\ binary populations and galaxy physical properties
(e.g., stellar mass and stellar ages), we have gained insight into the key
factors that influence the formation and evolution of compact objects, the
stars involved in the accreting binary phase, and their associated remnants
(e.g., millisecond pulsars and compact-object binaries).  However, a comprehensive
understanding of the formation and evolution of compact objects will require
empirical constraints on \xray\ binary populations in galaxies at a
variety of evolutionary stages.


\begin{table*}
\begin{center}
\caption{Properties of Early-Type Galaxy Sample}
\begin{tabular}{lcccccccccccc}
\hline\hline
  &  \multicolumn{2}{c}{ {\sc $K$-band Position}} & & &  &  & &  & & &  \\
  &  \multicolumn{2}{c}{\rule{1.0in}{0.01in}} & & &  &  & &  &  & & \\
  & $\alpha_{\rm J2000}$ & $\delta_{\rm J2000}$ & $D$ & $a$ & $b$ & PA & $r_{\rm exclude}$ & $\log L_{K}^{\prime}$ &  & & {\sc Age} & SFR \\
 \multicolumn{1}{c}{{\sc Galaxy}} & (hr) & (deg) & (Mpc) & (arcmin) & (arcmin) & (deg) & (arcsec) & (\lksol) & $f_{K, {\rm exclude}}$ & $S_N$ & (Gyr) & ($M_\odot$ yr$^{-1}$)  \\
\multicolumn{1}{c}{(1)} & (2) & (3) & (4) & (5) & (6) & (7) & (8) & (9) & (10) & (11) & (12) & (13) \\
\hline\hline

      NGC3115     & 10 05 14.0 & $-$07 43 07 &  {\phn9.68}$\pm$0.41  &   2.7 &   1.1 &  315  &    20 & 10.95$\pm$0.04 &   0.46 &     2.3$\pm$0.5 &    8.5$^{+1.0}_{-1.0}$$^a$       &      $<$0.004                 \\
      NGC3379                             & 10 47 49.6 & +12 34 53 & 10.57$\pm$0.55  &   3.2 &   2.7 &  338  &    10 & 10.89$\pm$0.05 &   0.25 &     1.3$\pm$0.7 &    8.6$^{+1.0}_{-1.0}$$^a$       &      $<$0.004                 \\
      NGC3384                             & 10 48 16.9 & +12 37 45 & 11.59$\pm$0.77  &   3.5 &   1.8 &  321  &    20 & 10.78$\pm$0.06 &   0.47 &     1.1$\pm$0.5 &    3.2$^{+1.5}_{-1.0}$$^b$ &      $<$0.009  \vspace{0.05in}\\
\hline
\end{tabular}
\end{center}
NOTE.---Col.(1): Target name. Col.(2) and (3): Right ascention and declination,
respectively, based on $K$-band data from the 2MASS large galaxy atlas (Jarrett
\etal\ 2003).  Col.(4): Distance and 1$\sigma$ error as determined by Tonry
\etal\ (2001).  Col.(5)--(7): $K$-band galaxy major axis, minor axis, and
position angle, respectively.  Col.(8): Radius of a small circular region,
centered on the galactic nucleus, that was excluded from further analyses due
to both the presence of a bright optical background from the galactic stellar
field and significant \xray\ source confusion. Col.(9): Logarithm of the total
galaxy-wide $K$-band luminosity and 1$\sigma$ error.  The 1$\sigma$ errors are
dominated by the errors on the distances (see Column~(4)).  These luminosities
have {\it not} been corrected for the excluded region specified in Column~(8).
Col.(10): Fraction of the total galaxy-wide $K$-band luminosity contained
within the excluded region.  The final $L_K$ value used in most calculations
throughout this paper can be calculated as $L_K = L_{K}^{\prime} \times
(1-f_{\rm K, exclude})$, where $L_{K}^{\prime}$ is the value provided in
Column~(9).  For our sample, $L_K$ and $L_{K}^{\prime}$ can be converted to
stellar mass using $M_\star/L_K \approx 0.66 M_\odot/L_{K,\odot}$, appropriate
for all three galaxies.  Col.(11): Globular cluster specific frequency $S_N$
from Harris~(1991). Col.(12): Stellar age estimates as provided by
$^a$S{\'a}nchez-Bl{\'a}zquez \etal\ (2006) or $^b$McDermid \etal\ (2006).
Col.(13): Integrated galaxy-wide star-formation rates inferred from \spitzer\
from Temi \etal\ (2009) or Shapiro \etal\ (2010).
\end{table*}


In the Milky Way, it has been observed that low mass \xray\ binaries (LMXBs)
are the dominant luminous accreting binary population, and are collectively a
factor of $\approx$10 times more luminous than the high mass \xray\ binaries
(e.g., Grimm \etal\ 2002).  With the advent of sub-arcsecond imaging with
\chandra, it has now been shown that LMXB populations are prevalent in other
nearby galaxies (e.g., Fabbiano~2006 and references therein). \chandra\ studies
of \xray\ binaries in local galaxies have found that the galaxy-wide integrated
\xray\ power from LMXBs is strongly correlated with galaxy
stellar mass ($M_\star$; e.g., O'Sullivan \etal\ 2001; Colbert \etal\ 2004;
Gilfanov~2004; Kim \& Fabbiano~2010; Lehmer \etal\ 2010; Boroson \etal\ 2011;
Zhang \etal\ 2012).  However, in early-type galaxies, the RMS scatter in the
$L_{\rm X}$--$M_\star$ relation is found to be $\approx$40\%, which is a
factor of $\simgt$3--5 times larger than typical measurement error and a factor
of $\approx$2 times larger than that expected from statistical variations due
to small number statistics (see Gilfanov~2004).  This indicates that additional
factors (e.g., stellar age and globular cluster LMXB systems) aside from
$M_\star$ alone influence the prevalence of LMXBs in galaxies.

Population synthesis models predict that $L_{\rm X}$ per $M_\star$ for LMXBs in
galactic fields should decrease with increasing stellar age.  This is primarily
due to a decline in the typical masses of LMXB donor stars, and \xray\
luminosities, with increasing stellar age (e.g., Belczynski \etal\ 2002, 2008;
Fragos \etal\ 2008, 2013a, 2013b).  Initial observational tests of these models
have yielded conflicting results.  In support of the population synthesis
predictions, Kim \& Fabbiano (2010) studied samples of seven young
(\hbox{$\approx$2--5~Gyr}) and seven old (\hbox{$\approx$8--12~Gyr}) nearby
early-type galaxies using \chandra\ observations that probe the $L_{\rm X}
\simgt 10^{38}$~\lum\ end of the \xray\ binary luminosity function (XLF).
Their observations showed that the fraction of LMXBs that were more luminous
than $\approx$$5 \times 10^{38}$~\lum\ is a factor of $\approx$2 times higher
($\approx$3$\sigma$) for the young early-type galaxy sample.  Further support
for the population synthesis expectation came from Lehmer \etal\ (2007), who
utilized \xray\ stacking analyses in the Extended \chandra\ Deep Field-South
survey to find that optically faint early-type galaxies (\hbox{$L_B
\approx$~$10^{9.5}$--$10^{10}$~\lbsol}), a population expected to have \xray\
emission dominated by LMXBs, have mean LMXB emission that is a factor of
\hbox{$\approx$3--8} times larger at $z \approx 0.4$ ($\approx$4~Gyr ago) than
today.  
More recently, however, Zhang \etal\ (2012) constructed detailed LMXB XLFs for
a sample of 20 early-type galaxies spanning luminosity-weighted stellar ages of
\hbox{1.6--17~Gyr} and found an {\it increase} in the $K$-band luminosity
normalized LMXB XLF normalization with increasing stellar age.   The same
conclusion can be drawn from the total LMXB $L_{\rm X}/L_K$ versus stellar age
diagrams constructed from the sample of 30 early-type galaxies studied by
Boroson \etal\ (2011) (see $\S$6 and Figure~9$a$).  These studies seemingly
contradict previous results and the population synthesis expectation.  

Although the above studies and theoretical expectations seem to show apparent
conflict, many of the differences between studies can be explained by
inhomogeneities in the galaxy samples and LMXB populations being studied.
Although there has been substantial efforts on dynamic modeling of globular
clusters (GCs), published population synthesis modeling exclusively follows the
evolution of LMXBs produced in primordial binaries formed in galactic fields
(hereafter, field LMXBs) and does not make simultaneous predictions for LMXBs
formed in GCs.  The GC LMXBs form efficiently through dynamical interactions of
multiple bodies (e.g., Fabian \etal\ 1975; Hills~1976; Verbunt~1987), and
their formation is expected to be tied to the dynamical and chemical properties
of the host GC with little relation to the host galaxy stellar population age.
In GCs, LMXBs are expected to form continuously as stars interact dynamically.
While in galactic fields, LMXB formation and emission is expected to be
directly connected to stellar evolution timescales of the secondary stars,
lagging Gyrs behind previous star-formation episodes.

Detailed studies of LMXB populations of local early-type galaxies have been
substantially biased towards massive galaxies, which tend to have rich GC
systems (e.g., Bekki \etal\ 2006) and LMXB populations dominated by GC LMXBs
and not field LMXBs (e.g., Irwin~2005).  Going forward, it will be critical to
test population synthesis models by studying specifically the {\it field LMXB}
populations in a sample of passive galaxies with stellar populations that cover
a large range of stellar ages.  The combination of deep {\itshape Chandra} and
{\it Hubble Space Telescope} (\hst) imaging offers a means for identifying and
separating LMXBs that are formed in galactic fields and GCs (e.g., Voss \etal\
2009).
At present, such a study has yet to be performed, primarily because the
necessary deep {\itshape Chandra} and {\itshape HST} observations of early-type
galaxies with relatively young to intermediate stellar ages ($\approx$3--5~Gyr)
had not yet been undertaken.

To remedy the above limitation, we have conducted a pilot program consisting of
deep \chandra\ and \hst\ observations of the nearby early-type galaxy NGC~3384,
which has a luminosity-weighted stellar age of $\approx$$3.2 \pm 1.5$~Gyr
(e.g., McDermid \etal\ 2006; S{\'a}nchez-Bl{\'a}zquez \etal\ 2007).  In this
paper, we utilize our new \chandra\ and \hst\ data to construct the field LMXB
XLF for the ``young'' stellar population in NGC~3384.  We compare the field
LMXB XLF of NGC~3384 with those derived using deep archival \chandra\ and \hst\
data of two ``old'' early-type galaxies, NGC~3115 and NGC~3379, which have
luminosity-weighted stellar ages of $8.5 \pm 1.0$ and $8.6 \pm 1.0$~Gyr
(S{\'a}nchez-Bl{\'a}zquez \etal\ 2006), respectively.  The stellar ages for the
whole sample are based on results from simple stellar population model fitting
to both absorption lines and spectral energy distribution shapes.  In this
process, absorption-line strengths were all calibrated to the Lick/IDS system
(e.g., Trager \etal\ 1998) and stellar ages were determined by interpolating
between multidimensional model grids of absorption line indices and ages.  All
three galaxies have been shown to have smooth spatial distributions of
luminosity-weighted stellar ages consistent with no significant age variations
across the galactic extents (Norris \etal\ 2006; S{\'a}nchez-Bl{\'a}zquez
\etal\ 2007).

Throughout this paper, we make use of three \xray\ bandpasses:
\hbox{0.5--2~keV} (soft band [SB]), \hbox{2--8~keV} (hard band [HB]), and
\hbox{0.5--8~keV} (full band [FB]).  Values of $H_0$ = 70~\hbox{km s$^{-1}$
Mpc$^{-1}$}, $\Omega_{\rm M}$ = 0.3, and $\Omega_{\Lambda}$ = 0.7 are adopted
throughout this paper (e.g., Spergel \etal\ 2003).

\begin{table*}
\begin{center}
\caption{\chandra\ Advanced CCD Imaging Spectrometer (ACS) Observation Log}
\begin{tabular}{lcccccc}
\hline\hline
  &  & {\sc Obs. Start} & {\sc Exposure Time}$^a$ & {\sc Flaring}$^b$ & {\sc Flaring Time}$^b$ & \\
\multicolumn{1}{c}{\sc Galaxy} & {\sc Obs. ID} & (UT) & (ks) & {\sc Intervals} & (ks) & {\sc Obs. Mode}$^c$ \\
\hline\hline
                   NGC~3115\ldots\ldots\ldots\ldots\ldots\ldots\ldots\ldots &           2040 &        2001 Jun 14, 10:31    &      \phn\phn35.3  &    1 &        1.6    &   F   \\
                                                                            &          11268 &        2010 Jan 27, 00:42    &      \phn\phn40.6  &    0 &        \ldots &  VF   \\
                                                                            &          12095 &        2010 Jan 29, 14:29    &      \phn\phn75.7  &    0 &        \ldots &  VF   \\
                                                                            &          13817 &        2012 Jan 18, 14:08    &         \phn172.0  &    0 &        \ldots &  VF   \\
                                                                            &          13819 &        2012 Jan 26, 22:21    &      \phn\phn70.0  &    2 &        5.5    &  VF   \\
                                                                            &          13820$^d$ &    2012 Jan 31, 14:55    &         \phn184.2  &    0 &        \ldots &  VF   \\
                                                                            &          13821 &        2012 Feb \phn3, 09:39 &         \phn157.5  &    1 &        0.5    &  VF   \\
                                                                            &          13822 &        2012 Jan 21, 08:54    &         \phn143.0  &    2 &        17.2   &  VF   \\
                                                                            &          14383 &        2012 Apr \phn4, 02:37 &         \phn119.0  &    1 &        0.5    &  VF   \\
                                                                            &          14384 &        2012 Apr \phn6, 17:51 &      \phn\phn69.7  &    0 &        \ldots &  VF   \\
                                                                            &          14419 &        2012 Apr \phn5, 18:49 &      \phn\phn46.3  &    0 &        \ldots &  VF   \\
                                                                            &         Merged &        \ldots                &       {\bf 1113.3} &    7 &     25.3   & \ldots    \\
                                               NGC~3379\ldots\ldots\dotfill &           1587 &        2001 Feb 13, 01:33    &      \phn\phn30.5  &    2 &        1.0    &   F   \\
                                                                            &           7073$^d$ &    2006 Jan \phn1, 05:50 &      \phn\phn81.6  &    3 &        2.5    &   F   \\
                                                                            &           7074 &        2006 Apr \phn9, 10:34 &      \phn\phn69.1  &    0 &        \ldots &  VF   \\
                                                                            &           7075 &        2006 Jul \phn3, 11:04 &      \phn\phn82.1  &    2 &        1.0    &  VF   \\
                                                                            &           7076 &        2007 Jan 10, 01:51    &      \phn\phn69.2  &    0 &        \ldots &  VF   \\
                                                                            &         Merged &        \ldots                &    \phn{\bf 333.5} &    7 &       4.5    &  \ldots     \\
                                               NGC~3384\ldots\ldots\dotfill &           4692 &        2004 Oct 19, 18:57    &   \phn\phn\phn9.9  &    0 &        \ldots &  VF   \\
                                                                            &          11782 &        2010 Jan 19, 11:31    &      \phn\phn28.2  &    1 &        0.5    & F   \\
                                                                            &          13829$^d$ &    2012 Jun 23, 11:07    &      \phn\phn94.8  &    0 &        \ldots & VF   \\
                                                                            &         Merged &        \ldots                &    \phn{\bf 132.9} &    1 &       \phn0.5    & \ldots     \\
\hline
\end{tabular}
\end{center}
Note.---Links to the data sets in this table have been provided in the electronic edition. \\
$^a$ All observations were continuous. These times have been corrected for removed data that was affected by high background; see $\S$~3.1.\\
$^b$ Number of flaring intervals and their combined duration.  These intervals were rejected from further analyses. \\
$^c$ The observing mode (F=Faint mode; VF=Very Faint mode).\\
$^d$ Indicates Obs.~ID by which all other observations are reprojected to for alignment purposes.  This Obs.~ID was chosen for reprojection as it had the longest initial exposure time, before flaring intervals were removed.\\
\end{table*}


\begin{table*}
\begin{center}
\caption{\hst\ Advanced Camera for Surveys (ACS) Observation Log}
\begin{tabular}{lccccccc}
\hline\hline
  &   &  &  &  & \multicolumn{3}{c}{\sc Line Dither Pattern}  \\
  &   &  &  &  & \multicolumn{3}{c}{\rule{1.8in}{0.01in}}  \\
  &    &  &  \multicolumn{2}{c}{{\sc Exposure Time} (s)} & \multicolumn{2}{c}{$N_{\rm pts}$} &  \\
  & {\sc Field}   & {\sc Obs. Start}  &  \multicolumn{2}{c}{\rule{1.2in}{0.01in}} & \multicolumn{2}{c}{\rule{1.0in}{0.01in}} & {\sc Spacing}  \\
\multicolumn{1}{c}{\sc Galaxy} & {\sc Number} & (UT) & (F475W)  & (F850LP)  & (F475W)  & (F850LP)  & (arcsec) \\
\hline\hline
                   NGC~3115\ldots\ldots\ldots\ldots\ldots\ldots\ldots\ldots &              1 &        2012 Mar 7, 02:49  &   824  &  1170  &   2  &  3 & 3.011 \\
                                                                            &              2 &        2012 Mar 7, 04:25  &   824  &  1170  &   2  &  3 & 3.011 \\
                                                                            &              3 &        2012 Mar 7, 06:14  &   722  &  1137  &   2  &  4 & 3.011 \\
                                                                            &              4 &        2012 Mar 7, 08:06  &   824  &  1170  &   2  &  3 & 3.011 \\
                                                                            &              5 &        2012 Mar 9, 17:05  &   824  &  1170  &   2  &  3 & 3.011 \\
                                                                            &              6 &        2012 Mar 10, 00:07  &   824  &  1170  &   2  &  3 & 3.011 \\ 
                                               NGC~3379\ldots\ldots\dotfill &              2 &        2010 Mar 25, 04:08  &   760  &  1285  &   2  &  3 & 0.145 \\
                                                                            &              3 &        2010 Apr 01, 00:38  &   754  &  1275  &   2  &  3 & 0.145 \\
                                                                            &              4 &        2010 Apr 01, 02:14  &   754  &  1275  &   2  &  3 & 0.145 \\
                                                                            &              5 &        2010 Apr 01, 03:50  &   754  &  1275  &   2  &  3 & 0.145 \\
                                                                            &              6 &        2010 Apr 01, 05:26  &   754  &  1275  &   2  &  3 & 0.145 \\
                                               NGC~3384\ldots\ldots\dotfill &              4 &        2012 Apr 10, 14:30  &   780  &  1305  &   2  &  3 & 2.994 \\
                                                                            &              5 &        2012 Apr 18, 08:59  &   800  &  1290  &   2  &  3 & 2.994 \\
                                                                            &              6 &        2012 Apr 14, 03:06  &   780  &  1305  &   2  &  3 & 2.994 \\
\hline
\end{tabular}
\end{center}
Note. --- Links to the data sets in this table have been provided in the electronic edition. \\
\end{table*}


%
\section{Early-Type Galaxy Sample Selection}
%

Our primary goal is to measure and compare field LMXB XLFs down to a relatively
faint X-ray luminosity limit for early-type galaxies spanning a wide range of
stellar ages.  We therefore required a sample of early-type galaxies with deep
\chandra\ observations that have corresponding \hst\ imaging across the whole
galactic extents.  

Using a base sample of 376 nearby early-type galaxies from Ellis \&
O'Sullivan~(2006), we searched the \chandra\ archive (from Cycle~12) for
observations that were sufficient to detect \xray\ point sources to a
highly-complete 0.5--8~keV depth of $L_{\rm X} \approx 10^{37}$~\lum, an
important limit above which a significant number of LMXBs can be detected (see
below) and a variety of LMXB donors and accretors are expected to contribute
(Fragos \etal\ 2008).  The Ellis \& O'Sullivan~(2006) sample was selected from
the Lyon-Meudon Extragalactic Data Archive (LEDA) using morphological type ($T
< 1.5$; \hbox{E--S0} Hubble types), distance ($V \le 9000$~km s$^{-1}$), and
apparent magnitude ($B_T \le 13.5$).  To probe effectively the stellar-age
dependent LMXB activity in early-type galaxies, we isolated galaxies that (1)
contain passive stellar populations with little ongoing star-formation as
measured by detailed \spitzer\ studies (e.g., Temi \etal\ 2009; Shapiro \etal\
2010), (2) have small values of GC specific frequency, $S_N \simlt 3$, (where
$S_N \equiv N_{\rm GC} 10^{0.4(M_V+15)}$, $N_{\rm GC}$ is the number of GCs
within the galaxy, and $M_V$ is the absolute $V$-band magnitude of the galaxy),
to study optimally both field and GC LMXBs, but without GC LMXBs dominating
(e.g., Kim \& Fabbiano~2004; Irwin~2005; Juett \etal\ 2005; Boroson \etal\
2011; Zhang \etal\ 2012), (3) have distances $D \simlt 15$~Mpc, as to avoid
serious confusion issues, and (4) have $K$-band luminosities (a proxy for
stellar mass) in a relatively small range \hbox{$L_{K} \approx$~2--10~$\times
10^{10}$~\lksol}.\footnote{$K$-band luminosities were computed using data from
the Two Micron All Sky Survey (2MASS) large galaxy atlas (Jarrett \etal\ 2003)
and $K$-band extended source catalogs.  The 2MASS extended source catalog is
available online via http://tdc-www.harvard.edu/catalogs/tmxsc.html.}  The
final $K$-band luminosity criterion removes large variations in LMXB
populations that might arise as a result of the $L_{\rm X}$--$M_\star$
relation, ensures that a sufficient, statistically meaningful number of LMXBs
(\hbox{$\simgt$50--100}) per galaxy are detected above $L_{\rm X} \approx
10^{37}$~\lum, and minimizes contributions from hot \xray\ gas that are
prevalent in more massive early-type galaxies (David \etal\ 2006).

From the criteria above, we found that three galaxies, Cen~A, NGC~3115, and
NGC~3379 had sufficient \chandra\ coverage to study in detail LMXB XLFs;
however, Cen~A is problematic for our program, since it has highly uncertain
global stellar population ages due to large dust-lanes, shows signatures of
recent gas-rich merger activity (e.g., Kainulainen \etal\ 2009), and further
contains bright extended \xray\ structures (shells and a jet) making it
difficult to study the XRB populations (e.g., Voss \etal\ 2009).  By contrast,
NGC~3115 and 3379 have deep \chandra\ and \hst\ exposures and satisfy all four
selection criteria posed above and are therefore ideal candidates to include in
our program (see, e.g., Kim \etal\ 2006, 2009; Brassington \etal\ 2008, 2010).
These galaxies have been determined to have relatively old stellar populations
($\approx$8--10~Gyr based on multiple stellar age measurements) and therefore
these \chandra\ and \hst\ observations have the potential to provide key
observational constraints for ``old'' field LMXB populations.  

Using the remaining early-type galaxy sample that did not have deep \chandra\
data in the archive, we searched for ``young'' ($\simlt$5~Gyr) early-type
galaxy candidates to target with \chandra\ and \hst.  We found that NGC~3377
and 3384 were the nearest sources ($D < 15$~Mpc) that have relatively young
stellar populations and satisfy our selection criteria.  In Cycle~13, we
successfully proposed for joint \chandra\ and \hst\ observations of NGC~3384,
which we present for the first time in this paper.  Taken together, NGC~3115,
3379, and 3384 form a clean sample of passive $D < 15$~Mpc early-type galaxies
with stellar ages ranging from \hbox{$\approx$3--10~Gyr}.  The \chandra\ and
\hst\ observations of all three galaxies cover the entire $K$-band extents and
reach 0.5--8~keV limits of $\simlt$$10^{37}$~\lum.  The $K$-band luminosities
provide good proxies for the stellar masses $M_\star$ of the galaxies in our
study.  Adopting the prescription outlined in Appendix 2 of Bell \etal\
(2003), we utilized $B-V$ colors to estimate that $M_\star/L_K \approx 0.66 \;
M_{\odot}/L_{K,\odot}$ for all three galaxies in our sample.  We note that this
method has an uncertainty of \hbox{$\approx$0.1--0.2~dex} on the derived
$M_\star/L_K$ for a given $B-V$ color.  It is likely the case that the younger
galaxy NGC~3384 has a {\itshape true} $M_\star/L_K$ that is lower than those
for the two older galaxies NGC~3115 and 3379.  Hereafter, we make use of
scalings with $L_K$, and where appropriate, convert to $M_\star$ using a
constant $M_\star/L_K$ conversion factor for the whole sample.  In Table~1, we
summarize the properties of the galaxies that make up our sample and in $\S$~3
we describe our \chandra\ and \hst\ data and analyses.  

%
\section{Data Reduction}
%

As discussed in $\S$~2, we are comparing results from new \chandra\ and \hst\
observations of the ``young'' early-type galaxy NGC~3384 with
comparable-to-better quality archival data for ``old'' early-type galaxies
NGC~3115 and NGC~3379.  We note that there are published papers describing some
of the \chandra\ and \hst\ archival data available for NGC~3115 and NGC~3379
(e.g., Brassington \etal\ 2008, 2010, 2012; Wong \etal\ 2013); however, to
limit comparison errors, we analyze the \chandra\ and \hst\ data for
all three galaxies using the methodologies adopted in this paper.  In the
sections below, we describe our data reduction methods in detail.

%
%
\begin{figure*}
\figurenum{1}
\centerline{
\includegraphics[width=18cm]{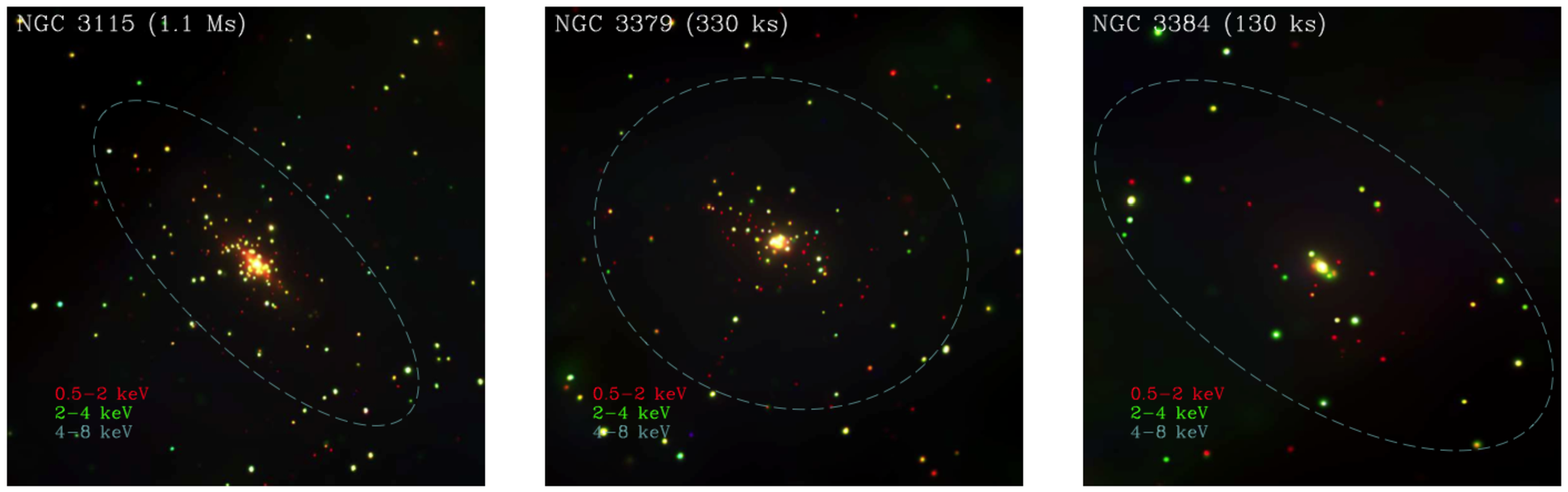}
}
\vspace{0.1in}
\caption{
Three-color \chandra\ images of the three early-type galaxies in our sample.
Each image was constructed from 0.5--2~keV ($red$), 2--4~keV ($green$), and
4--8~keV ($blue$) exposure-corrected adaptively smoothed images.  Dashed
ellipses represent the total $K$-band galaxy size and orientation as
described in Columns~(2)--(3) and (5)--(7) of Table 1.  We note that the
range of the \chandra\ image depths (see annotations) is large (0.1--1.1~Ms),
thus affecting
comparitive quality. 
}
\end{figure*}
%
%
\begin{figure*}
\figurenum{2}
\centerline{
\includegraphics[width=18cm]{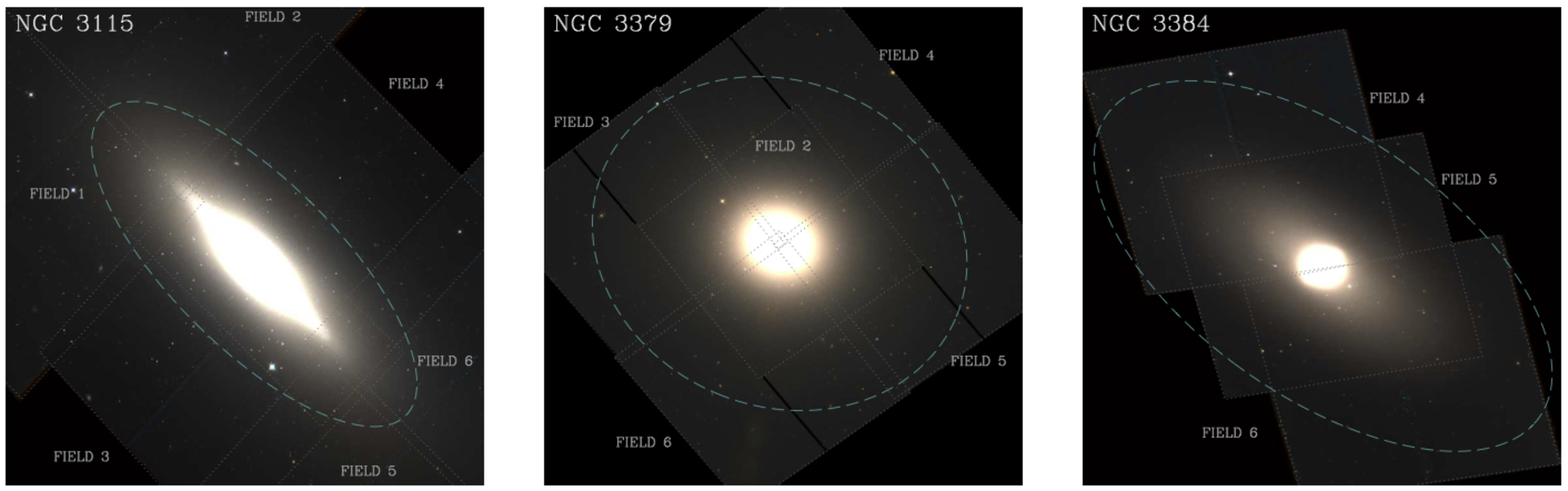}
}
\vspace{0.1in}
\caption{
Color \hst\ ACS mosaics of the three early-type galaxies in our sample.  Each
image was constructed using F850LP ($red$), (F850LP + F475W)/2 ($green$), and
F475W ($blue$) image mosaics.  Image sizes, orientations, and dashed ellipses
are the same as in Figure~1.  Each $202\arcsec \times 202\arcsec$ ACS field of view
has been outlined ({\it solid squares\/}) and the field name annotated.  A
description of our mosaicking procedure can be found in $\S$~3.2.
}
\end{figure*}

\subsection{Chandra Data Reduction and X-ray Catalog Creation}

The \chandra\ observations for all three galaxies were conducted in multiple
observations (hereafter, ObsIDs) using the S-array of the Advanced CCD Imaging
Spectrometer (ACIS-S; see Table~2 for full observation log).  The $K$-band
semi-major axes of the galaxies (Column~5 of Table~1) range from
2.7\arcmin--3.5\arcmin\ and therefore subtend an angular extent smaller than a
single ACIS-S chip ($\approx$$8\farcm5 \times 8\farcm5$).  As such, for all
\chandra\ observations, the galactic extents were almost entirely contained
within the ACIS-S3 chip for all observations.  For our data reduction, we made
use of {\ttfamily CIAO}~v.~4.5 with {\ttfamily CALDB}~v.~4.5.6.  We began by
reprocessing the pipeline produced events lists, bringing level~1 to level~2
using the script {\ttfamily chandra\_repro}.  The {\ttfamily chandra\_repro}
script runs a variety of {\ttfamily CIAO} tools that identify and remove events
from bad pixels and columns, and filter the events list to include only good
time intervals without significant flares and non-cosmic ray events
corresponding to the standard \asca\ grade set (\asca\ grades 0, 2, 3, 4, 6).

Using the reprocessed level~2 events lists for each ObsID, we generated
preliminary FB images and point-spread function (PSF) maps (using the tool
{\ttfamily mkpsfmap}) with a monochromatic energy of 1.497~keV and an encircled
counts fraction (ECF) set to 0.393.   For each ObsID, we created preliminary
source catalogs by searching the FB images with {\ttfamily wavdetect} (run
including our PSF map), which was set at a false-positive probability threshold
of $1 \times 10^{-5}$ and run over seven wavelet scales from 1--8 pixels (1,
$\sqrt{2}$, 2, 2$\sqrt{2}$, 4, 4$\sqrt{2}$, and 8).  To measure sensitively
whether any significant flares remained in our observations, we constructed
point-source-excluded \hbox{0.5--8~keV} background light curves for each ObsID
with 500~s time bins.  For each galaxy, we found \hbox{1--7} intervals across
all ObsIDs with flaring events of $\simgt$3~$\sigma$ above the nominal
background; these intervals were removed from further analyses (see Table~2).
In general, this resulted in the removal of short intervals; however, for
NGC~3115 $\approx$25~ks of data were excluded.  The resulting cumulative
exposure times for NGC~3115, 3379, and 3384 were 1,113~ks, 333~ks, and 133~ks,
respectively.

For each galaxy and ObsID, we utilized the preliminary source catalogs based on
{\ttfamily wavdetect} positions to register each good-time-interval screened
aspect solution and events list to the frame with the longest exposure time
(see Table~2); this was achieved using {\ttfamily CIAO} tools {\ttfamily
reproject\_aspect} and {\ttfamily reproject\_events}, respectively.  The
resulting astrometric reprojections gave nearly negligible linear translations
($<$0.38~pixels), rotations ($<$0.1~deg), and stretches ($<$0.06\% of the pixel
size) for all ObsIDs.  We created merged events lists for each galaxy using the
reprojected events lists and the {\ttfamily CIAO} script {\ttfamily
merge\_obs}.

Using the individual ObsID and merged events lists, we constructed images for
each galaxy in the FB, SB, and HB.  For each of the three bands, we created
corresponding exposure maps following the basic procedure outlined in $\S$3.2
of Hornschemeier \etal\ (2001); these maps were normalized to the effective
exposures of sources located at the aim points.  This procedure takes into
account the effects of vignetting, gaps between the CCDs, bad column and pixel
filtering, and the spatially-dependent degredation of the ACIS optical blocking
filter.  A photon index of $\Gamma=1.4$, the slope of the extragalactic cosmic
\hbox{X-ray} background in the full band (e.g., Hickox \& Markevitch~2006), was
assumed in creating the exposure maps.  

For a given galaxy, we created a {\it candidate source catalog} by first
searching each image (i.e., all ObsIDs and merged images in all three bands)
using {\ttfamily wavdetect} at a liberal false-positive probability threshold
of $1 \times 10^{-5}$ over the $\sqrt{2}$ sequence (see above).  We searched
all ObsIDs to isolate the locations of transient source candidates that may
have been diluted by background and thus not detected by {\ttfamily wavdetect}
in the merged image.  We then merged the {\ttfamily wavdetect} catalogs
together by cross-matching the point-source catalogs.  For a given galaxy, we
cross-matched the catalogs using matching radii of 1\farcs5, 2\farcs5, and
3\farcs5 for sources offset by $\simlt$2\arcmin, 2\arcmin--6\arcmin, and
$\simgt$6\arcmin, respectively, from the exposure-weighted average aim-point
location.  We inspected all the images and candidate source regions by eye to
see if any additional source candidates were missed.  In particular, in the
crowded nuclear regions of the elliptical galaxies, low-flux point sources may
not be picked up by {\ttfamily wavdetect} due to the complex backgrounds.  As a
result of our inspection, we identified six additional candidate sources,
including 3, 1, and 2 sources in the nuclear regions of NGC~3115, 3379, and
3384, respectively, which we added to our candidate source catalogs.

We performed detailed photometry on our candidate source catalogs using the
{\ttfamily ACIS EXTRACT} (hereafter, {\ttfamily AE}; Broos \etal\ 2012)
point-source analysis software and the {\ttfamily wavdetect}-based
positions.\footnote{The {\ttfamily ACIS EXTRACT} software is available on the
WWW at http://www.astro.psu.edu/xray/docs/TARA/ae\_users\_guide.html} The \ae\
software contains algorithms that allow for appropriate computation of source
properties when multiple observations with different roll angles and/or aim
points are being combined and analyzed (see discussion below for further
details).  \ae\ also uses complex algorithms for accurately computing the
photometry of sources in crowded regions like those found in the nuclear
regions of our early-type galaxies.  The improved photometry provided by \ae\
allows us to evaluate clearly the significances of the point sources in each
candidate source catalog and prune our source lists to include reliable
sources.  
For a source to be considered detected, it must contain $s$ counts within an
aperture representing $\approx$90\% of the point-source encircled-energy
fraction (EEF) satisfying the following binomial probability criterion:
\begin{equation}
P(x \ge s) = \displaystyle\sum\limits_{x=s}^{n} \frac{n!}{x! (n-x)!} p^x (1 -
p)^{n-x} \le P_{\rm thresh},
\end{equation}
where $n \equiv s + b_{\rm ext}$ and $p \equiv 1/(1+b_{\rm ext}/b_{\rm src})$.
Here $b_{\rm ext}$ is the total number of background counts extracted from a
large region outside of the point source (while masking out regions from other
\xray\ detected sources) that was used to obtain an estimate of the local
background count rate.  The quantity $b_{\rm src}$ is the estimated number of
background counts within the source extraction region, which was measured by
rescaling $b_{\rm ext}$ to the area of the source aperture (the inter-quartile
range of $b_{\rm ext}/b_{\rm src} \approx$~24--118).  We adopted a threshold
value of $P_{\rm thresh} = 0.004$, below which we considered a source candidate
to be a valid X-ray point source and unlikely to be a fluctuation of the
background.  Our choice was motivated by the Xue \etal\ (2011) analysis of the
4~Ms \chandra\ Deep Field-South survey, in which the use of $P_{\rm thresh} =
0.004$ maximized the number of valid sources that had optical/near-IR
counterparts without introducing a significant number of false sources.  

For each galaxy, we constructed {\it main catalogs} by iteratively pruning our
candidate source lists to include only sources with $P \le P_{\rm thresh}$ in
at least one of the three bandpasses.  In this process, we first computed $P$
using {\it only} the merged photometry,\footnote{Had we chosen to consider $P$
computed from both the merged observations and each of the individual
observations, additional faint variable sources would be included in our main
catalogs.  However, the deliberate inclusion of such sources would result in 
biasing our XLFs upwards for increasing numbers of ObsIDs.  To some
extent, bright variable sources may have a small effect on our results, if
the sources are detected in the merged images. } we then re-extracted source
and background counts from the remaining sources, and repeated the process
until no further sources were removed.  The three galaxies required 1--3
iterations before converging.  

We note that our multi-stage procedure for identifying a highly reliable list
of source candidates will not result in a complete selection of all real
sources in each image having binomial probabilities $\le P_{\rm thresh}$, since
the initial {\ttfamily wavdetect} selection has more complex source-detection
criteria (see Freeman \etal\ 2002 for details) than the simple criterion given
in equation~1.  Such a caveat is important when considering completeness of our
\xray\ catalogs.  In $\S$~5 we discuss our methods for correcting for these
completeness issues when computing LMXB XLFs.

\subsection{HST Data Reduction and Optical Catalog Creation}

Each of the three galaxies was observed with \hst\ using the Advanced Camera
for Surveys (ACS) with the F475W and F850LP filters; hereafter the $g_{475}$
and $z_{850}$ bands, respectively.  The ACS field of view subtends an area of
$\approx$$202\arcsec \times 202\arcsec$; therefore the full mapping of each
galaxy required multiple pointings.  For a given galaxy, each ACS pointing
pattern was carefully chosen to cover the entire optical extent of the galaxy
and achieve higher sensitivity near the galactic center.  High sensitivity in
the central regions of the galaxies is crucial for effectively identifying GCs
and unrelated background candidates that fall within the bright stellar
emission near the galactic centers.  Details regarding the ACS observations are
logged in Table~3, and in Figure~2 we show mosaicked ACS images of the three
galaxies in our sample with each of the $202\arcsec \times 202\arcsec$
observational fields highlighted.  

For each galaxy, we first constructed a mosaicked image in the $g_{475}$ band
by aligning each of the images to a chosen reference observation and
``drizzling'' (Fruchter \& Hook~2002) the images to the common frame.  For
NGC~3115, we first aligned ACS fields~3 and 4 (see annotations in Fig.~2) to
create a large central field that contained some overlapping areas common to
the remaining ACS fields~1, 2, 5, and 6; these fields were subsequently aligned
to the large central field.  For NGC~3379 and NGC~3384, we aligned all
observations to the central images (i.e., ACS fields~2 and 5 for NGC~3379 and
NGC~3384, respectively).  We performed field alignments using the {\ttfamily
Pyraf} tools {\ttfamily Tweakreg} and {\ttfamily Tweakback}, which were
available through the {\ttfamily drizzlepac}~v.4.3.0\footnote{See
http://drizzlepac.stsci.edu/ for details on {\ttfamily drizzlepac}.} software
package.  {\ttfamily Tweakreg} and {\ttfamily Tweakback} identify sources that
are common to each ACS image and a chosen reference image (here the galactic
centers) and update the image headers once an astrometric solution is found.
Given the small overlaps between images, we implemented only small shifts in
right ascension and declination to align our images.  In general,
\hbox{$\approx$4--200} bright sources per image were used in the field
alignments, and the resulting shifts and RMS residuals to the fits were
$\simlt$10~ACS pixels ($\simlt$0\farcs5) and $\simlt$0.1~pixels
($\simlt$0\farcs005), respectively.  After aligning the ACS fields, we
constructed $g_{475}$ mosaicked images using the {\ttfamily astrodrizzle} tool
within {\ttfamily drizzlepac}.  The {\ttfamily astrodrizzle} procedure uses the
aligned, flat-field calibrated and charge-transfer efficiency (CTE) corrected
images to create a distortion-corrected mosaicked image with bad pixels and
cosmic rays removed.  

We repeated the above ACS image mosaicking procedure for the $z_{850}$ band
data, with the exception that all ACS observations of a given galaxy were
aligned to the full $g_{475}$ mosaicked image.  The resulting two-colored
mosaicked images are displayed in Figure~2 with the $K$-band galactic extents
overlaid. 

To construct source catalogs, we first searched each image mosaic using
{\ttfamily SExtractor} (Bertin \& Arnouts~1996).  We chose the following
{\ttfamily SExtractor} parameters: minimum detection area of 5~pixels;
detection and photometry analysis thresholds of 3$\sigma$; background grid of
$16 \times 16$ pixels with a filtering Gaussian of 2.5~pixels full width at
half-maximum (FWHM); and 32 deblending sub-thresholds.  The {\ttfamily
SExtractor} catalogs include some non-negligible contributions from background
fluctuations and unrejected cosmic rays (e.g., in chip gap regions where
dithering makes cosmic-ray rejection difficult).   For each galaxy, we
constructed a master optical catalog, in which we required that sources be
detected in both the $g_{475}$ and $z_{850}$ bands, using a matching radius of
0\farcs5.  

Within the central 10\arcsec--20\arcsec\ of the galactic centers, the stellar
field backgrounds are very high, particularly in the $z_{850}$ band, and our
{\ttfamily SExtractor} searching does not yield source detections in these
regions.  Since we are unable to identify optical counterparts to \xray\
sources (e.g., GCs and background sources) in these regions, we reject the
\xray\ sources in these regions from consideration when computing field LMXB
XLFs.  As we will describe in $\S$5, these regions are also subject to
significant \xray\ point source confusion, which supports our choice to reject
these regions.  The sizes of these rejected regions have been tabulated in
Column~(8) of Table~1, and the fraction of the $K$-band luminosity contained
within these regions, as measured by the $K$-band data, are provided in
Column~(10) in Table~1.  Hereafter, we utilize $K$-band luminosities that have
been corrected for the exclusion of these regions.

%
%
\begin{figure*}
\figurenum{3}
\centerline{
\includegraphics[width=20cm]{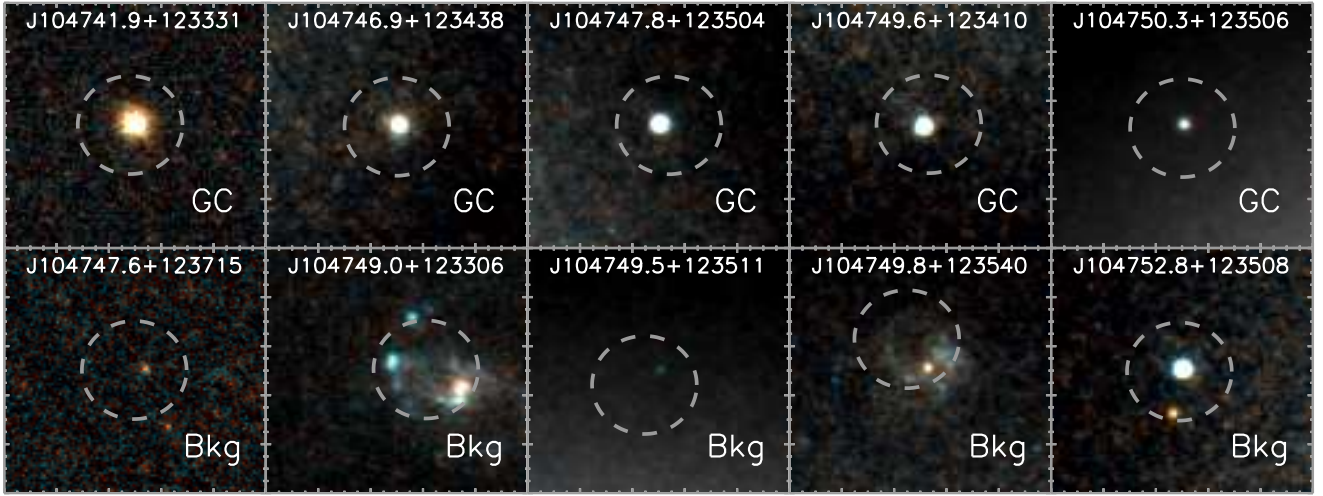}
}
\vspace{0.1in}
\caption{
Sample ACS image postage stamps of \xray\ detected GC ({\it top panels\/}) and
unrelated background source ({\it bottom panels\/}) candidates in NGC~3379.
Our optical source counterpart classification criteria are described in $\S$~4.
Each postage stamp is centered on the optical source position and spans a
$2\farcs5 \times 2\farcs5$ region.  For scaling purposes, we have included a
circle with 1\arcsec\ radius centered on the location of the \xray\ source.
}
\end{figure*}

\begin{table}
\begin{center}
\caption{Summary of X-ray Source Classifications in Galaxy Footprints}
\begin{tabular}{lccccc}
\hline\hline
 \multicolumn{1}{c}{Galaxy}  & $N_{\rm tot}$ & $N_{\rm nuc}$ & $N_{\rm GC}^{\rm LMXB}$ & $N_{\rm bkg}$ & $N_{\rm field}^{\rm LMXB}$ \\
 \multicolumn{1}{c}{(1)}  & (2) & (3) & (4) & (5) & (6) \\
\hline\hline
            NGC 3115\ldots\ldots\ldots\ldots\ldots & 153 &  46 &  25 &   9 &  73 \\
            NGC 3379\dotfill & 151 &  17 &  10 &  18 & 106 \\
            NGC 3384\dotfill &  65 &  18 &   3 &  14 &  30 \\
\hline
\end{tabular}
\end{center}
NOTE.---Col.(1): Galaxy name. Col.(2) Total number of X-ray sources detected within the galaxy footprint, including the nuclear regions. Col.(3) Number of nuclear sources that were excluded from XLF computations (see $\S\S$~3.2 and 5.2 for explanation and justification). After excluding the nuclear sources, Col.(4)--(6) provide the number of \xray\ detected sources classified as GC, unrelated background, and field LMXB candidates, respectively.
\end{table}


\subsection{Astrometric Alignment of \chandra\ and \hst\ Frames and Catalog Summaries}

We further refined the absolute astrometry of our \chandra\ and \hst\ data
products and catalogs by registering them to either the Eighth Sloan Digital
Sky Survey (SDSS) Data Release (DR8; Aihara \etal\ 2011) frame (NGC~3379 and
3384) or the United States Naval Observatory CCD Astrograph Catalog (UCAC2;
Zacharias \etal\ 2000) frame when SDSS DR8 data were unavailable (NGC~3115).
SDSS DR8 was astrometrically calibrated to UCAC2 frame; however, the SDSS DR8
reaches a much deeper limiting $r$-band magnitude of $r_{\rm lim} \approx 22.5$
(50\% complete), compared to $r_{\rm lim} \approx 16$ for UCAC2, and is
therefore our preferred optical reference frame.

To align our \hst\ frames with reference optical frames (i.e., SDSS DR8 or
UCAC2), we first cross-matched the \hst\ master optical catalogs of each galaxy
with the reference catalog using a 1\farcs5 matching radius.  We then applied
median astrometric shifts in right ascension and declination (total offsets
ranged from 0\farcs25--0\farcs56 for the three galaxies) to the \hst\ images
and source catalogs to bring them into alignment with the reference frame.  To
align the \chandra\ catalogs and data products, we matched the \chandra\ main
catalogs of each galaxy to their corresponding astrometry-corrected \hst\
master optical catalogs using a matching radius of 1\farcs5.  Again, simple
median shifts in right ascension and declination (offsets ranged from
0\farcs17--0\farcs26 for the three galaxies) were applied to the \chandra\ data
products and catalogs to bring them into alignment with the \hst\ and reference
optical frames (by extension).  

The final \hst\ and \chandra\ image and catalog registrations have a 1$\sigma$
error of $\approx$0\farcs12.  Our further analyses are focused on sources that
lie within the $K$-band extents of the galaxies outside the crowded nuclear
regions (see Table~1).  Within these galactic footprints, our main \chandra\
catalogs (see $\S$3.1) contain 107, 134, and 47 \xray\ sources for NGC~3115,
3379, and 3384, respectively, with the faintest FB detected sources reaching
flux levels of 5.9, 7.9, and $16 \times 10^{-17}$~\flux, respectively.  

The optical limits of our \hst\ catalogs vary substantially across the galactic
footprints due to large background gradients from the galactic stellar fields.
Within the galactic footprints (after excluding the inner regions specified in
Table~1), our optical catalogs contain 582, 819, and 1125 total sources, which
imply source densities of 64.8, 30.2, and 57.9 arcmin$^{-2}$ for NGC~3115,
3379, and 3384, respectively.  As we will point out later, the factor of
$\approx$2 times lower source density for NGC~3379 is also evident in the
background \xray\ number counts and is may be due to cosmic variance on the
(small) size scale of the galactic footprints (see $\S$5.3).  Within the
galactic footprints, we find that the $g_{475}$-band limiting depths of our
catalogs is deepest in the innermost regions (outside the excluded radii given
in Column~(8) in Table~1) where the overlapping \hst\ coverage generally
contains a factor of $\sim$2 times deeper imaging than the galactic outskirts.
Despite the deeper imaging in the central regions, we find that the
$z_{850}$-band limiting depths are shallowest in the inner regions of the
galaxy due to the enhanced background from the host galaxies in these regions.
Specifically, in the inner regions of the galaxies, we find 5$\sigma$ limiting
depths of $g_{475}^{\rm lim} \approx (26.0, 26.0, 26.3)$ and $z_{850}^{\rm lim}
\approx (24.5, 24.5, 24.7)$ for NGC~3115, 3379, and 3384, respectively.  In the
outer regions, the respective 5$\sigma$ limiting depths are $g_{475}^{\rm lim}
\approx (25.6, 25.6, 26.0)$ and $z_{850}^{\rm lim} \approx (24.9, 24.9, 25.1)$.

%
\section{X-ray Source Classification}
%

For each galaxy, we cross-correlated our astrometry-corrected \chandra\ main
catalogs with our \hst\ ACS optical master catalogs to identify likely
counterparts to a subset of the \xray\ point sources.  An optical source was
considered to be a likely counterpart if its position was located within the
95\% confidence radius $r_{95}$ of the \xray\ source.  The size of $r_{95}$
depends on both net source counts and positional offset from the
exposure-weighted mean aim point; these dependencies were calculated following
equation~(12) of Kim \etal\ (2007).  In total, we found 34, 28, and 17 optical
counterparts to the \xray\ point sources that were within the galactic extents
of NGC~3115, 3379, and 3384, respectively; these represent 32\%, 21\%, and 36\%
of the \xray\ sources, respectively.  For each galaxy, we estimated the total
number of false matches expected, given our matching criteria.  To do this, we
first estimated the optical source density $\rho_{\rm opt}$ as a function of
angular distance from the galactic centers.  The expected number of false
matches was then calculated as $N_{\rm false} \approx \Sigma_i$ $\rho_{{\rm
opt},i} \times \Sigma_j A_{ij}$, where $\rho_{{\rm opt},i}$ is the density of
optical sources in the $i$th annulus and $A_{ij}$ is the area enclosed by the
95\% confidence positional error circle of the $j$th source residing in the $i$th
annulus.  We estimate that $N_{\rm false} \approx 1.1$, 2.4, and 1.3 for
NGC~3115, 3379, and 3384, respectively.  These values imply that $\approx$6\%
of the \xray\ sources with optical counterparts could be false matches, which
will not have a material effect on our results.

We expect that optical counterparts to the X-ray sources will fall in one of
two broad categories: LMXBs in GCs (expected to have $z_{850}
\approx$~18--24~mag, safely above the sensitivity limits in $\S$3.3) and
unrelated background sources.  Hereafter, we define unrelated background
sources to comprise all \xray\ sources that are not intrinsically related to
the galaxies being studied here.  This category therefore includes foreground
Galactic stars, as well as background AGN and galaxies.  Given that the most
optically luminous counterparts to field LMXBs are expected to be normal stars
with $g_{475}$ and $z_{850}$ magnitudes well below our detection thresholds
(e.g., luminous red giants will have $z_{850} \simgt$~28--31~mag at the
distances to our galaxies), we do not expect to detect optical counterparts to
individual field LMXBs.  We therefore consider \xray\ sources without optical
counterparts to be candidate field LMXBs.

%
%
\begin{figure*}
\figurenum{4}
\centerline{
\includegraphics[width=18cm]{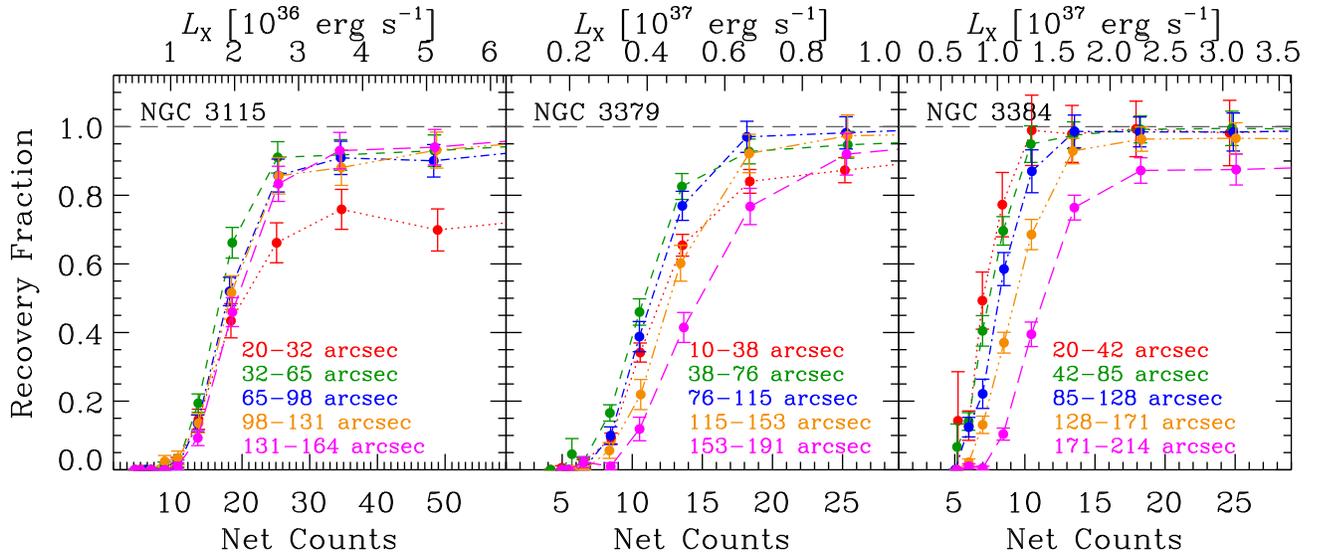}
}
\vspace{0.1in}
\caption{
Fraction of recovered simulated sources as a function of net source counts for
our three early-type galaxies (filled circles with 1$\sigma$ error bars).  A
simulated source was considered ``recovered'' if (1) the total number of
on-source counts satisfied the binomial probability selection criterion of
equation 1, (2) the source was detected by {\ttfamily wavdetect} using a
false-positive probablity threshold of 10$^{-5}$, and (3) the source had net
counts that were within 3$\sigma$ of the input counts (required to identify
confused sources). Each color represents a different offset range from the
nuclear position, as annotated in each plot.  Our best-fit models have been
plotted as dashed curves.  For a complete description of our simulations, see
$\S$~4.2.
}
\end{figure*}

For \xray\ sources with optical counterparts, we performed basic source
classifications using their optical colors and light-profile information.  Optical
sources were classified as likely GCs if they had (1) colors in the range of
\hbox{$0.6 \le g_{475}-z_{850} \le 1.6$}; (2) absolute magnitudes (based on the
distances to each galaxy) in the range of $-12.5 \le M_z \le -6.5$; (3)
extended light profiles in either the $g_{475}$ or $z_{850}$ bandpasses,
characterized as having {\ttfamily SExtractor} stellarity parameters {\ttfamily
CLASS\_STAR}~$\le$~0.9 or aperture magnitude differences $m(r_1)-m(r_2) > 0.4$,
where apertures consist of circles with radii $r_1 = 0\farcs25$ and $r_2 = 0\farcs5$;
and (4) light profiles in both the $g_{475}$ and $z_{850}$ bands that were not
too extended to be GCs, defined as $m(r_1)-m(r_2) \le 0.9$.  \xray\ sources
with optical counterparts that did not meet these criteria were classified as
unrelated background objects.

In Figure~3, we show ten random examples of \xray\ sources in NGC~3379
classified as GC and unrelated background sources.   We visually inspected
images similar to those shown in Figure~3 for all optical counterparts in all
three galaxies.  We found that our adopted selection criteria yielded
reasonably well segregated samples of GCs and unrelated background sources with
few sources where the classifications are obviously incorrect.  Furthermore,
the inferred \hbox{0.5--8~keV} luminosities of the GC counterparts were all
less than $L_{\rm X} \approx 7 \times 10^{38}$~\lum, which is well within the
expected range for GC LMXBs.  We note that it is likely that a small minority
of sources will have incorrect source classifications (e.g., faint sources with
poor constraints on spatial properties); however, we do not expect these
misclassifications to have any material effects on our conclusions.  We
also note that our ability to identify and separate GC LMXBs from field LMXBs
is dependent on the completeness of our GC catalogs.  Using the full \hst\
catalogs, we constructed observed optical luminosity functions (i.e., number of
detected sources per magnitude as a function of absolute magnitude $M_z$) for
sources detected within the galactic footprints.  For all three galaxies, we
found steadily rising number counts with decreasing $M_z$ down to limits of
$M_z^{\rm lim} \approx -5.0$ to $-5.5$ before incompleteness began to affect
source detection.  As discussed above, GCs were selected to to have $-12.5 \le
M_z \le -6.5$, boundaries that are well above the highly-complete limits.  Our
observed GC optical luminosity functions are consistent with those of the Milky
Way, which to first order can be described as a Gaussian with peak luminosity
at $M_z \approx -7.9$~mag and $\sigma \approx 1.3$~mag (Harris ~1991).  We
therefore expect our GC selections to be highly complete.

In Table~4, we summarize our optical source categorizations for the \xray\
sources.  As discussed above, we have excluded from consideration sources that
were within small regions of the galactic centers due to the very bright
stellar fields in these regions (see $\S$3.2).  The number of excluded sources
has been provided in Column~(3) of Table~4.  Within the acceptable regions of
the galaxy extents, we find 30--106 field LMXBs and 3--25 GC LMXBs per galaxy.
In the sections below, we describe our procedure for generating LMXB XLF
populations and present our resulting XLFs.

%
\section{X-ray Luminosity Function Computation and Fitting}
%

We chose to compute XLFs in the FB (0.5--8~keV) to make direct
comparisons with recent XLF studies of early-type galaxies (e.g., Zhang \etal\
2012).  Use of the FB for identifying \xray\ point sources is appropriate for
our early-type galaxy sample, since these galaxies have negligible
contamination from hot gas (see, e.g., Fig.~1) and little absorption at
$\simlt$1~keV from cold interstellar gas and dust.  To compute XLFs for each of
our galaxies, we made use of the FB-detected sources in the main \chandra\
catalogs presented in $\S$3.1.  Spatial variations in \xray\ sensitivity, due
to variations in local backgrounds, PSFs, effective exposures (e.g., chip gaps
and bad pixels and columns), and source crowding, combined with incompleteness
can have significant effects on the shape of the XLFs at \xray\ luminosities
that are within a factor of $\approx$10 of the detection limits.  These effects
need to be accounted for to properly compute the XLFs.  
Our approach, described in detail below, is to fit the {\it observed} XLFs with
models of the {\it intrinsic} LMXB XLFs convolved with completeness functions
derived from simulations. 

\subsection{X-ray Background and Sensitivity Maps}

Our first step in computing completeness functions was to construct FB
background and sensitivity maps for each of the three galaxies in our sample.
Background maps were constructed by first masking out all \xray\ point sources
within the merged FB images using circular masking regions of size 1.1 times
the radius that encloses 99\% of the PSF.  These masking regions were computed
and utilized by \ae\ in the process of computing local background estimates for
the point sources.  After masking out the point sources, we filled the masked
regions with Poisson noise that was characteristic of the local backgrounds as
computed by \ae.  The background maps enabled us to create sensitivity maps,
which contain pixel-by-pixel estimates of the minimum number of counts $s_{\rm
min}$ that a hypothetical source would need to be detected given the
local background.  To compute $s_{\rm min}$ at each location in a
given image, it is necessary to estimate local values of $b_{\rm src}$ and
$b_{\rm ext}$ (via Eqn.~(1)) that are similar to those that would be estimated
following the \ae\ approach used for the \xray\ detected sources.
Unfortunately, the \ae\ procedure for extracting these values at every location
in the image is computationally prohibitive, since source photometry is
performed on an observation-by-observation basis using computationally
intensive ray-tracing algorithms to generate local PSFs and approximations to
the 90\% EEF PSF contours.  To overcome this issue, we estimated $b_{\rm src}$
by extracting background counts from the background maps using circular
apertures with sizes that encompass the 90\% EEF for a point source.
Appropriate values of $b_{\rm ext}$ were estimated at each location (pixel) by
extracting the maximum value of $b_{\rm ext}$ from the main \chandra\ catalogs
for sources with off-axis angles $\theta = \theta_p \pm 0\farcm25$, where
$\theta_p$ is the pixel off-axis angle.  Given values of $b_{\rm src}$ and
$b_{\rm ext}$, we numerically solved the relation $P(x \ge s_{\rm min}) =
0.004$ (i.e., Eqn.~(1)) to obtain $s_{\rm min}$ at each image location.  In
this manner, we constructed a spatial sensitivity map consisting of $s_{\rm
min}$ values.

%
%
\begin{figure}
\figurenum{5}
\centerline{
\includegraphics[width=9.5cm]{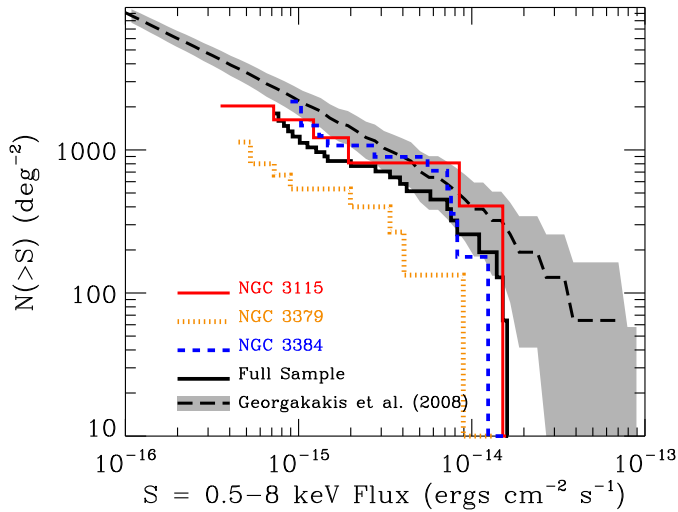}
}
\caption{
0.5--8~keV number counts measurement for \xray\ sources within the galactic
footprints of our early-type galaxies that were optically classified as
unrelated background objects ({\it see annotations\/}).  For comparison, we show
the expected number counts of unrelated background objects ({\it dotted
curve\/}), and 1$\sigma$ error envelope ({\it grey region\/}), based on number
counts from extragalactic \xray\ surveys (Georgakakis \etal\ 2008).  This
comparison indicates that we are able to identify the majority of unrelated
background \xray\ sources using our methods.  The deficit of number counts in
NGC~3379 may be due to cosmic variance (see discussion in $\S$5.3).
}
\end{figure}

\subsection{X-ray Completeness Functions}

The above maps allow us to determine spatially whether a source would have been
detected for a given number of total counts.  However, these maps alone cannot
directly be used to assess source completeness and encapsulate subtleties
related to our complex selection process (i.e., {\ttfamily wavdetect} plus \ae\
pruning) and source confusion in crowded regions.  To assess better
source detection completeness as a function of net source counts for each
\chandra\ data set, we made use of simulated data sets and a Monte Carlo procedure that
replicated the source detection and pruning processes that were used to create
our source catalogs.

We began by generating 800 mock FB images for each galaxy with 35 \xray\
sources artificially added to the original FB images.  Ideally, we would have
generated more mock images with only 1 \xray\ source implanted in each mock
image, so that the confusion properties would not be changed by the addition of
sources.  However, we found empirically that the addition of 35 sources per image both
avoids having any non-negligible effect on the source confusion properties of the original
images and allows for 35 times more efficient execution of the simulation
process.  For each \xray\ source added to a given mock image, we assigned random
right ascension and declination values that were probabilistically offset from
the galaxy nucleus in accordance with the offset distribution of the main
\chandra\ catalog.  Each \xray\ source was given a random number of counts between
$\approx$1--100~counts.  The counts from each source were added to the
simulated image canvases using the {\ttfamily marx}\footnote{See
http://space.mit.edu/cxc/marx/ for {\ttfamily marx} simulator details}
(version~4.5) ray-tracing code.  In this procedure, the PSF of each source was
modeled assuming a roll angle and aim point that were selected from one of the
galaxy ObsIDs (Table~2).  Selection of the roll angle and aim point for each
source was done probabilistically with the probability of selection being
directly proportional to the exposure time of each of the observations.  

%
%
\begin{figure*}
\figurenum{6}
\centerline{
\includegraphics[width=18cm]{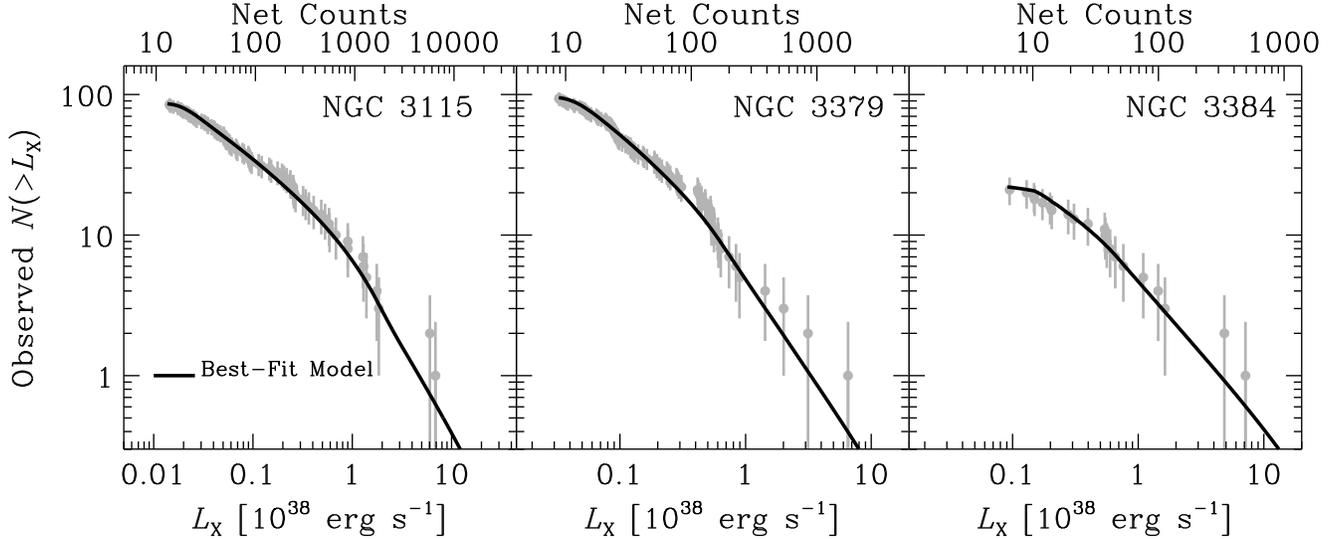}
}
\vspace{0.1in}
\caption{
Cumulative observed X-ray luminosity functions ($N(>L_{\rm X})$; {\it thick
black curves\/}) for the GC plus field LMXB populations of the three early-type
galaxies in our sample.  Each observed XLF ({\it gray circles with 1$\sigma$
Poisson error bars\/}) includes \xray\ detected point sources within the extent
of the galaxy after excluding unrelated background sources and sources within
the central nuclear regions (see text).  The XLFs were fit using an intrinsic
XLF model convolved with our completeness functions following the procedure
described in $\S$5.4.
}
\end{figure*}

For each galaxy, we performed circular aperture photometry on the 28,000 mock
\xray\ sources (i.e., 35 sources in each of the 800 mock images) using a
circular aperture with a radius encompassing $\approx$90\% of the
encircled energy.  The number of counts measured for each source was
then compared with the number of counts needed for a source detection from the
sensitivity maps to see if it satisfied the detection criterion adopted in
equation~(1).  We then searched each of the 800 mock images using {\ttfamily
wavdetect} at a false-positive probability threshold of $10^{-5}$ and made
{\ttfamily wavdetect} source catalogs for each image.  We considered a
simulated source to be ``recovered'' if (1) it was detected in the {\ttfamily
wavdetect} runs, (2) the total number of counts exceeded the value required for
a detection based on the sensitivity maps, and (3) the net counts derived were
within 3$\sigma$ of the known input counts.  Criterion~3 was adopted to account
for source confusion, since sources that are significantly confused with
brighter sources in the original image will have an excess of source counts due
to the unrelated source.  We consider such sources to not be recovered.

Our simulations revealed that the central $\approx$10\arcsec--15\arcsec\ of the
three early-type galaxies suffer catastrophic source confusion due to a high
density of bright X-ray sources in the nuclear regions.   As discussed in
$\S$3.2, these regions were considered problematic from an optical perspective,
since the bright emission associated with the stellar fields made it impossible
to identify potential GC and unrelated background candidates.  These regions
are problematic from the \xray\ perspective because (1) it is not possible to
recover accurately the true \xray\ luminosity distributions of the sources in
these regions; (2) the X-ray point sources may be formed via dynamical
interactions in the high stellar density environments (e.g., Voss \& Gilfanov
2007; Zhang \etal\ 2011), which are a different formation channel from the {\it
field} LMXBs that we are interested in studying here; and (3) among the X-ray
point sources of a given galaxy may be a low-luminosity AGN that could
contaminate the LMXB XLFs at the high-luminosity ends.  These considerations
further support our choice to exclude these regions from our XLF computations
below.

In Figure~4, we show the recovery fraction as a function of net source counts
in radial bins for our three early-type galaxies.  This plot was generated
using circular annuli including only sources that were within the elliptical
regions defined in Table~1; for the reasons discussed above, the exclusion
region defined in Column~(8) of Table~1 has not been shown.  For NGC~3115 and
3379, we find that source recovery becomes more complete moving from the
innermost annulus to the second innermost annulus.  This is due to confusion
becoming less severe at larger radii.  The completeness again deteriorates
going to even larger radii as the size of the PSF increases somewhat with
radial distance (N.B., the exposure-weighted \chandra\ aimpoints are all within
$\approx$5\arcsec\ of the galaxy centers).  For NGC~3384, confusion is not a
strong issue for the innermost annulus; therefore, the completeness simply
declines with radius.

\subsection{Unrelated Background X-ray Sources}

One of the key strategies in our XLF computations is to identify directly and
eliminate from consideration \xray\ point sources that are associated with
unrelated background objects.  Direct optical identification of background
\xray\ sources is preferred over statistical methods, in which a background
component is simply subtracted from the total XLFs, since the galaxies subtend
only small solid angles on the sky ($\simlt$0.01~deg$^2$) and are subject to
large variations in their background populations due to cosmic variance (e.g.,
Moretti \etal\ 2009).  To test the efficacy of our background identifications,
we constructed the \xray\ number counts ($\log N$-$\log S$; the number of
\xray\ sources per unit area $N$ with \xray\ flux brighter than $S$) relations
for each of the three galaxies (and the combined sample) using the \xray\
sources classified as unrelated background objects (see $\S$4).  For a given
galaxy, we limited this calculation to a flux regime in which we could have
detected sources across $\approx$75\% of the galactic extent.

Figure~5 shows the $\log N$-$\log S$ for the unrelated background sources in
the early-type galaxy footprints (as well as the total galaxy sample) compared
with expectation based on extragalactic blank-field \xray\ surveys (e.g.,
Georgakakis \etal\ 2008).  It is apparent that the number counts for unrelated
background sources for NGC~3115 and 3384 are in good agreement with the
expectation, while the NGC~3379 number counts fall a factor of
\hbox{$\approx$2.5--3} times below the expectation across the full \xray\ flux
range.  We expect that our optical identification and classification process
will be most complete and accurate at bright fluxes ($f_{\rm 0.5-8~keV}
\simgt$~few~$\times 10^{-15}$~\flux) where unrelated background counterparts
are likely to be relatively unobscured AGN with bright optical counterparts
(e.g., Steffen \etal\ 2006; Xue \etal\ 2011).  Since the deficit of \xray\
sources in NGC~3379 is apparent at these high fluxes, it is likely that the
deficit is real.  As noted in $\S$3.3, NGC~3379 also has a factor of $\approx$2
deficit in the source density of detected optical sources, lending strong
support to this conclusion.  Given the areal coverage of NGC~3379, we would
expect cosmic variance at a level of $\approx$30--35\% (see Moretti \etal\
2009).  Therefore the factor of $\simgt$2 deficit for NGC~3379, if due to
cosmic variance, would represent a $\simgt$3$\sigma$ deviation from the
expected value (assuming the area-dependent cosmic variance distribution is
symmetric and Gaussian).  

The above number counts analysis suggests that the majority of unrelated \xray\
background sources are likely to be classified as such by our procedure.  In
our computations of LMXB XLFs, described below, we remove unrelated background
\xray\ sources and do not try to correct for remaining contributions from
unrecovered background \xray\ sources.  In the faint \xray\ flux regime, where
our recovery process is expected to be the least complete, background \xray\
sources not classified as such are greatly outnumbered by field LMXBs and
therefore unlikely to have material effects on our results.

\begin{table*}
\begin{center}
\caption{Best-Fit Parameters to X-ray Luminosity Functions (XLFs)}
\begin{tabular}{lcccccccccccc}
\hline\hline
   & \multicolumn{4}{c}{GC + Field (Total) LMXBs} & \multicolumn{4}{c}{Field LMXBs} & \multicolumn{4}{c}{GC LMXBs} \\
   & \multicolumn{4}{c}{\rule{1.8in}{0.01in}} & \multicolumn{4}{c}{\rule{1.8in}{0.01in}} & \multicolumn{4}{c}{\rule{1.8in}{0.01in}} \\
 \multicolumn{1}{c}{Galaxy}  & $K$ & $\alpha_1$ & $\alpha_2$ & $L_b$ & $K$ & $\alpha_1$ & $\alpha_2$ & $L_b$ &   $K$ & $\alpha_1$ & $\alpha_2$ & $L_b$ \\
\hline\hline
            NGC 3115\ldots\ldots\ldots\ldots &                       6.8$\pm$1.8 &                       1.5$\pm$0.1 &                       2.2$\pm$0.7 &                     1.76$\pm$0.02 &                       4.6$\pm$1.2 &                       1.5$\pm$0.1 &                            \ldots & 1.853$^{+0.391}_{-0.004}$$^\star$ &                       2.8$\pm$1.2 &                       1.3$\pm$0.2 &                       1.7$\pm$0.6 &     1.76$\pm$0.19 \\
            NGC 3379\dotfill &       9.4$\pm$3.2 &       1.6$\pm$0.1 &       2.3$\pm$0.4 &     0.60$\pm$0.19 &       7.1$\pm$2.1 &       1.7$\pm$0.1 &       3.3$\pm$1.1 &     0.84$\pm$0.17 &       4.6$\pm$5.1 &       0.6$\pm$0.4 &       1.7$\pm$0.4 &                     0.41$\pm$0.26 \\
            NGC 3384\dotfill &       6.1$\pm$3.1 &       1.3$\pm$0.4 &       2.0$\pm$0.4 &     0.68$\pm$0.23 &       6.1$\pm$3.1 &       1.3$\pm$0.4 &       2.0$\pm$0.4 &     0.68$\pm$0.23 &            \ldots &            \ldots &            \ldots &                            \ldots \\
\hline
\end{tabular}
\end{center}
NOTE---$K$ and $L_b$ are in units of ($10^{38}$~\lum)$^{-1}$ and $10^{38}$~\lum, respectively. All errors are 1$\sigma$ and are based on the fitting procedure described in $\S$5.4.  For NGC~3384, all FB detected LMXBs have been classified as field candidates; therefore the GC plus field and field only fit parameters are identical.\\
$^\star$The best-fit model to the field LMXB XLF for NGC~3115 was a single power-law with normalization $K$, slope $\alpha_1$, and high-luminosity cut at the quoted $L_b$ value.
\end{table*}


%
%
\begin{figure*}
\figurenum{7}
\centerline{
\includegraphics[width=9.0cm]{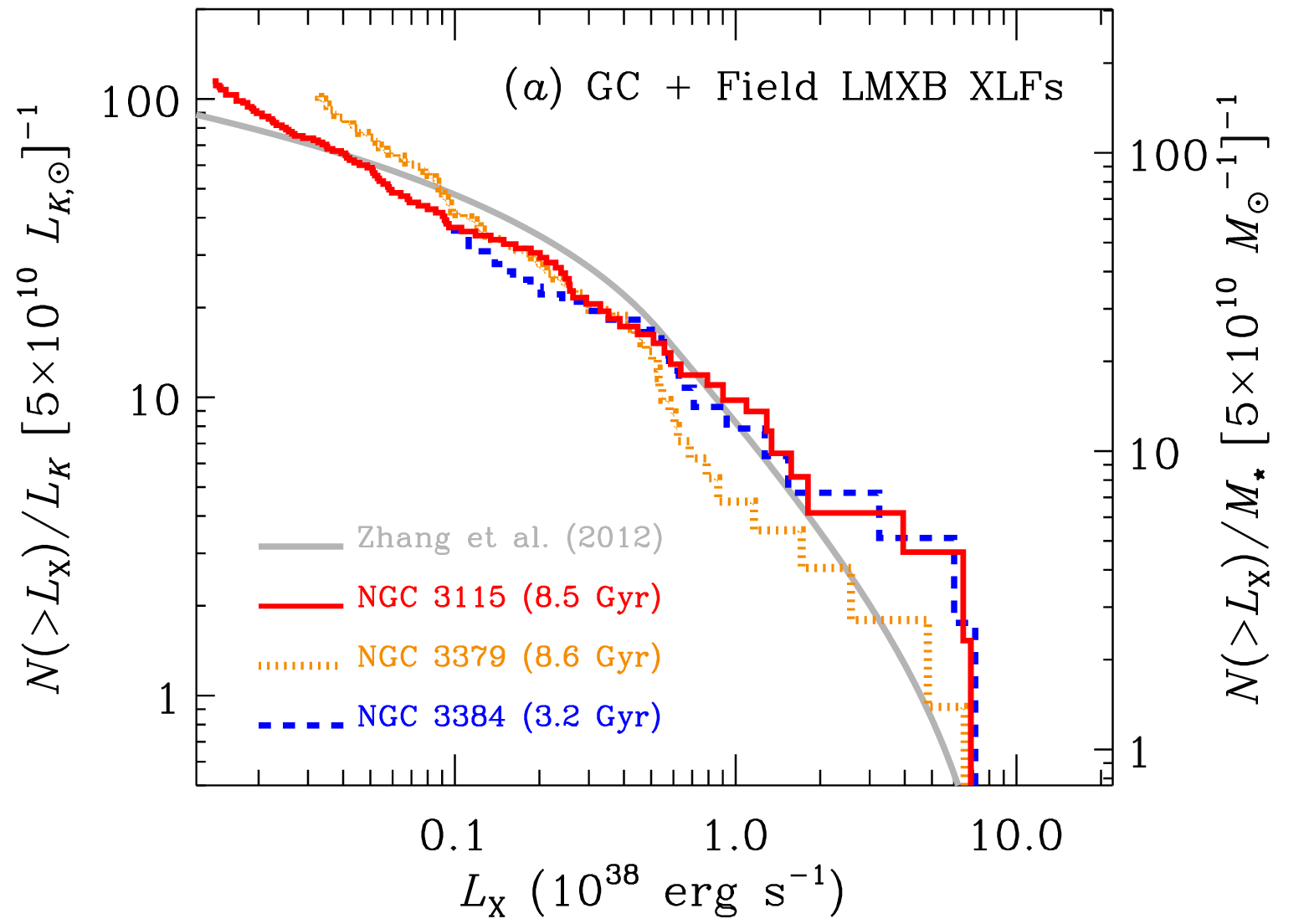}
\hfill
\includegraphics[width=9.0cm]{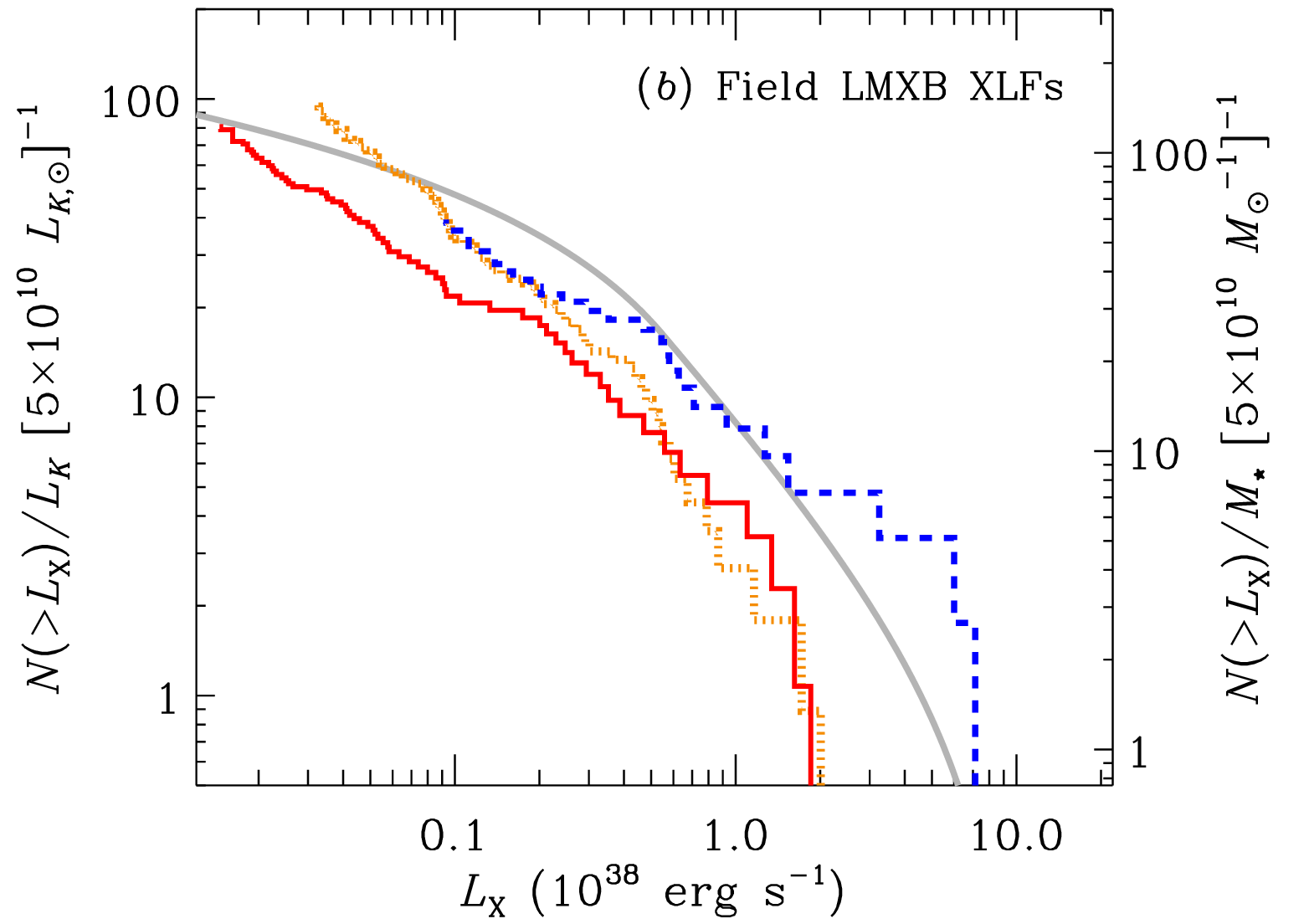}
}
\caption{
$K$-band--luminosity--normalized cumulative XLFs, $N(>L_{\rm X})/L_K$, for our
early-type galaxies in two LMXB population categories: ({\it a\/}) GC
plus field LMXBs and ({\it b\/}) field LMXBs.  Each XLF
has been summed down to an \xray\ limit that is roughly 50\% complete, and the
displayed XLFs have been corrected for completeness following the methods
described in $\S$5.4.  The three early-type galaxies have GC plus field (total) LMXB
XLFs consistent with each other and previous results (Zhang \etal\ 2012).
However, when looking at only the field LMXB XLF
contribution, we find that the young early-type galaxy NGC~3384 has a more
numerous and luminous field LMXB population than the old early-type galaxies
NGC~3115 and 3379 (see $\S$6 for more details).
}
\end{figure*}

\subsection{X-ray Luminosity Function Fitting}

For each galaxy, we constructed {\it observed XLFs} using the FB-detected
sources in the main \chandra\ catalogs that were within the galactic footprints
(excluding sources in the near vicinities of the galactic nuclei; see
Column~(8) of Table~1).  For each galaxy, we chose to construct and study the
XLFs for three LMXB population categories, including GC plus field (total) LMXBs, field
LMXBs, and GC LMXBs.  However, for NGC~3384, we found only three GC LMXB
candidates, which were all undetected in the FB, and therefore did not attempt
to construct their GC LMXB XLF.  In all cases, we excluded sources with optical
counterparts classified as unrelated background objects.  For illustration, we
have displayed the cumulative observed XLFs (i.e., $N(>L_{\rm X})$) for the GC
plus LMXB populations in Figure~6 ({\it gray points with error bars\/}).  

For each of the three LMXB sub-populations, we modeled {\it intrinsic} XLFs
using a broken power-law form that followed
\[dN/dL_{\rm X} = \left\{ 
\begin{array}{l r}
K (L_{\rm X}/L_{\rm ref})^{-\alpha_1}  & (L_{\rm X} \le L_{\rm b}) \\
K (L_{\rm b}/L_{\rm ref})^{\alpha_2 - \alpha_1}(L_{\rm X}/L_{\rm ref})^{-\alpha_2}, &
(L_{\rm X} > L_{\rm b}) \\ 
\end{array} \right. \]
where $K$ is the normalization at reference luminosity $L_{\rm ref} \equiv
10^{38}$~\lum, $\alpha_1$ is the faint-end slope, $L_{\rm b}$ is the break
luminosity, and $\alpha_2$ is the bright-end slope.  

We chose to model the observed XLFs using a forward-fitting approach, in which
we convolved an appropriately weighted completeness function with our model
XLFs before fitting to the observed XLFs.  As demonstrated in $\S$~5.2 and
Figure~4, our \xray\ completeness to sources depends upon their offsets from
the galactic centers.  For a given galaxy, we therefore statistically weighted
the contributions from model XLFs at each annulus according to the observed
distributions of \xray\ point sources to obtain the galaxy-wide weighted
completeness function $\xi(L_{\rm X})$.  Here we utilized only luminous \xray\
sources in regimes where our point source populations are highly complete at
all offsets from the galactic centers to determine the annular weightings.
Formally, we computed $\xi(L_{\rm X})$ using the following relation:
\begin{equation}
\xi(L_{\rm X}) = \sum_i f_{\rm recov, i}(L_{\rm X}) \times w_i,
\end{equation}
where $f_{\rm recov, i}(L_{\rm X})$ is the recovery-fraction curve for the
$i$th annular bin and $w_i$ is the fraction of total number of galaxy-wide
sources within the $i$th annuluar bin based on the observed point-source
distributions.

Using the {\ttfamily sherpa} modeling and fitting package available through
{\ttfamily ciao}, we modeled each observed XLF using a multiplicative model
\begin{equation}
dN/dL_{\rm X}({\rm observed}) = \xi(L_{\rm X}) dN/dL_{\rm X}({\rm model}).
\end{equation}
For a given subsample of sources, we constructed the observed $dN/dL_{\rm
X}({\rm observed})$ using a very small constant bin of $\delta L_{\rm X}$
spanning the minimum luminosity of the subsample to 10 times the maxmimum
luminosity.  Therefore the vast majority of bins contained zero sources up to a
maximum of three sources.  We evaluated the goodness of fit for our double
power-law models using the Cash statistic ({\ttfamily cstat}; Cash~1979).  We
tested to see if in any cases a single power-law provided a statistically
superior fit over the broken power-law, but were unable to find any such cases.
We did find that for the field LMXB XLF in NGC~3115, the best-fitting broken
power-law produced an effectively infinite high-luminosity slope ($\alpha_2$)
beyond the most luminous detected source.  For this subclass, we utilized a
single power-law with a high-luminosity cut-off (see Table~5).  In Figure~6, we
display the observed XLFs and best-fitting observed and intrinsic XLF models
for the GC plus field LMXB populations for our three early-type galaxies.  In
Table~5, we tabulate the best-fit parameters and 1$\sigma$ errors for fits to
the GC plus field LMXBs, field LMXBs, and GC LMXBs.

%
\section{Results}
%

In Figure~7, we show the $L_K$-normalized cumulative LMXB XLFs ($N(>L_{\rm
X})/L_K$) for our early-type galaxies for GC plus field and field LMXB
populations.  The XLFs displayed in Figure~7 have been corrected for
completeness, near the luminosity limits of each galaxy, using the completeness
corrections derived from our best-fitting XLFs (see $\S$5.4).  We find that the
GC plus field LMXB XLFs (Fig.~7$a$) for all three galaxies appear to be similar
in normalization and shape across $L_{\rm X} \approx 10^{37}$--$10^{39}$~\lum.
These estimates are in good agreement with the best-fit XLFs derived by Zhang
\etal\ (2012) for a sample of 20 massive early-type galaxies ({\it gray
curve\/}).  However, when we remove the contributions from GC LMXBs, we observe
significant galaxy-to-galaxy variations in the resulting field LMXB XLFs.  

In Figure~7$b$, we display the field LMXB XLF.  We find that the shapes and
normalizations of the field LMXB XLFs are similar for the two old early-type
galaxies, NGC~3115 and 3379, with some differences becoming apparent at $L_{\rm
X} \simlt$~\hbox{(2--4)}~$\times 10^{37}$~\lum.  For both of the old early-type
galaxies, we observe sharp declines in the field LMXB XLFs above $L_{\rm X}
\approx$~\hbox{(0.5--1)}~$\times 10^{38}$~\lum.  By contrast, the field LMXB
XLF for the young early-type galaxy NGC~3384 is generally elevated compared to
those of the old early-type galaxies and includes an extension to higher \xray\
luminosities ($L_{\rm X} \approx 7 \times 10^{38}$~\lum); however, as is
apparent from Figure~7$b$, the cumulative field XLF of NGC~3384 approaches that
of NGC~3379 at $L_{\rm X} \approx 3 \times 10^{37}$~\lum.  A similar excess of
luminous ($L_{\rm X} > 5\times 10^{38}$~\lum) LMXBs in young early-type
galaxies was previously observed by Kim \& Fabbiano (2010) and Zhang \etal\
(2012) using GC plus field LMXB populations.  Our analysis shows that this
excess may apparently be attributed to a strong excess in {\it field} LMXBs,
and the removal of GC LMXBs seems to amplify this excess of bright \xray\
sources.  This result clearly illustrates that \hst\ data are crucially needed
to identify directly LMXB counterparts and provide meaningful tests of
population synthesi models.  By selection, the galaxies in our sample have
relatively low $S_N$ ($S_N \simlt 3$), so the large elliptical galaxies that
have been studied well thus far will contain even larger contributions from GC
LMXBs.

%
%
\begin{figure}
\figurenum{8}
\centerline{
\includegraphics[width=9.0cm]{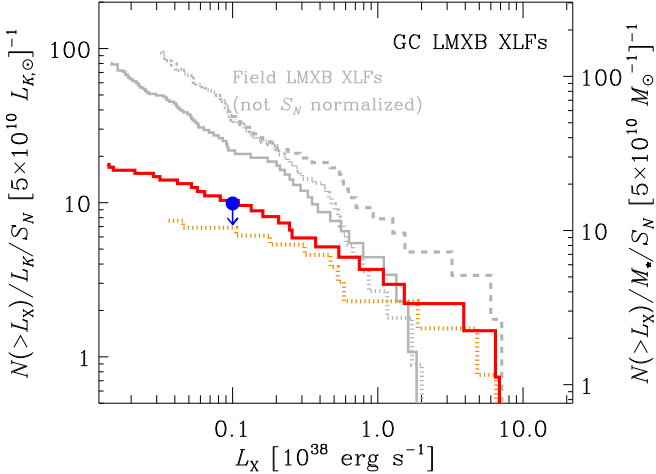}
}
\caption{
GC LMXB cumulative XLFs normalized by $L_K$ and $S_N$ ({\it colored curves\/})
with field LMXB cumulative XLFs, normalized by $L_K$ only, displayed for
comparison ({\it gray curves\/}).  Linestyles and colors follow the same scheme
as in Figure~7, with the exception of NGC~3384, for which we only obtain an
upper limit on the GC cumulative XLF at the detection limit ({\it filled blue
circle with arrow\/}).  We find that when normalized by $L_K$ and $S_N$,
the GC LMXB XLFs are all consistent with eachother in shape and normalization
and have overall slopes shallower than the field LMXB XLFs.
}
\end{figure}

In Figure~8, we display the GC LMXB cumulative XLFs normalized by $L_K$ and
$S_N$ \hbox{($N(>L_{\rm X})/L_K/S_N$)}.  We note that NGC~3384 had no GC LMXBs
detected in the FB; we therefore show only the 3$\sigma$ upper limit to its GC
XLF at the detection limit.  From Figure~8, we see that the GC LMXB XLFs for
NGC~3115 and 3379 appear to be broadly consistent in their shape and
normalization; the upper limit on the GC LMXB XLF for NGC~3384 is consistent
with the other galaxies.  For the purpose of comparing XLF shapes, we have
plotted in light gray $N(>L_{\rm X})/L_K$ (without the $S_N$ normalization) for
the field LMXBs.  We find that the GC LMXB XLFs of NGC~3115 and 3379 have
significantly shallower slopes and extend to higher \xray\ luminosities than
the field LMXB XLFs.  This result is consistent with previous results that
compared GC and field LMXB XLFs (e.g., Kim \etal\ 2009; Voss \etal\ 2009;
Paolillo \etal\ 2011).  

\begin{table}
\begin{center}
\caption{Integrated X-ray Luminosities}
\begin{tabular}{lcc}
\hline\hline
   & $\log L_{\rm X}$ & $\log L_{\rm X}/L_K$ \\
 \multicolumn{1}{c}{Galaxy}  & ($\log$ ergs s$^{-1}$) & ($\log$ ergs s$^{-1}$ $L_{K,\odot}^{-1}$) \\
\hline
\hline
\multicolumn{3}{c}{GC + Field (Total)} \\
\hline
            NGC 3115\ldots\ldots\ldots\ldots &  39.58 $+/-$ 0.19 &  28.90 $+/-$ 0.19 \\
            NGC 3379\dotfill &  39.51 $+/-$ 0.12 &  28.75 $+/-$ 0.12 \\
            NGC 3384\dotfill &  39.37 $+/-$ 0.26 &  28.87 $+/-$ 0.26 \\
\hline
\multicolumn{3}{c}{Field} \\
\hline
            NGC 3115\dotfill &  39.08 $+/-$ 0.10 &  28.40 $+/-$ 0.10 \\
            NGC 3379\dotfill &  39.35 $+/-$ 0.07 &  28.58 $+/-$ 0.07 \\
            NGC 3384\dotfill &  39.37 $+/-$ 0.26 &  28.87 $+/-$ 0.26 \\
\hline
\multicolumn{3}{c}{GC} \\
\hline
            NGC 3115\dotfill &  39.52 $+/-$ 0.34 &  28.84 $+/-$ 0.34 \\
            NGC 3379\dotfill &  39.17 $+/-$ 0.50 &  28.40 $+/-$ 0.50 \\
\hline
\end{tabular}
\end{center}
\end{table}


%
%
\begin{figure*}
\figurenum{9}
\centerline{
\includegraphics[height=7cm]{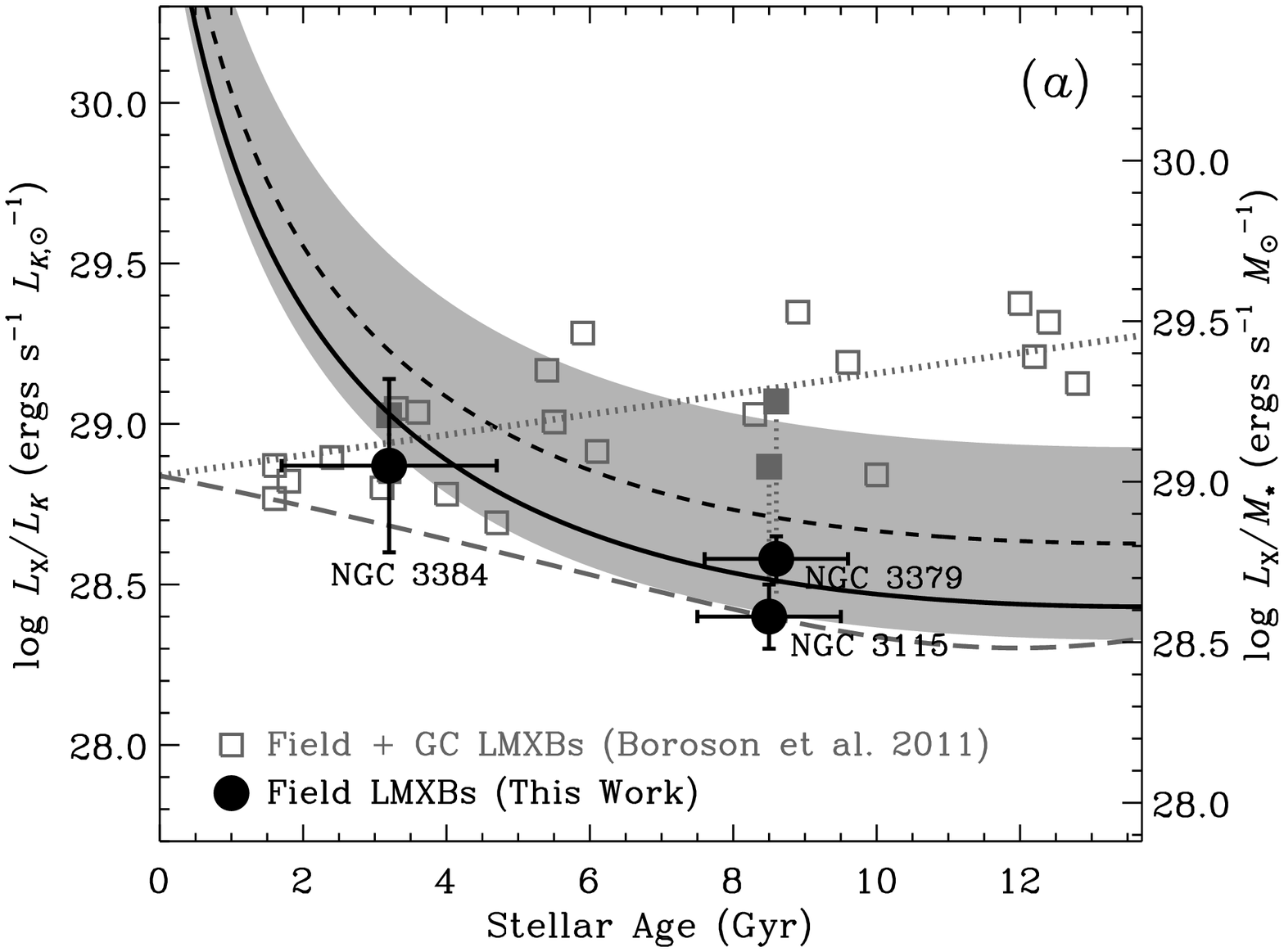}
\hfill
\includegraphics[height=7cm]{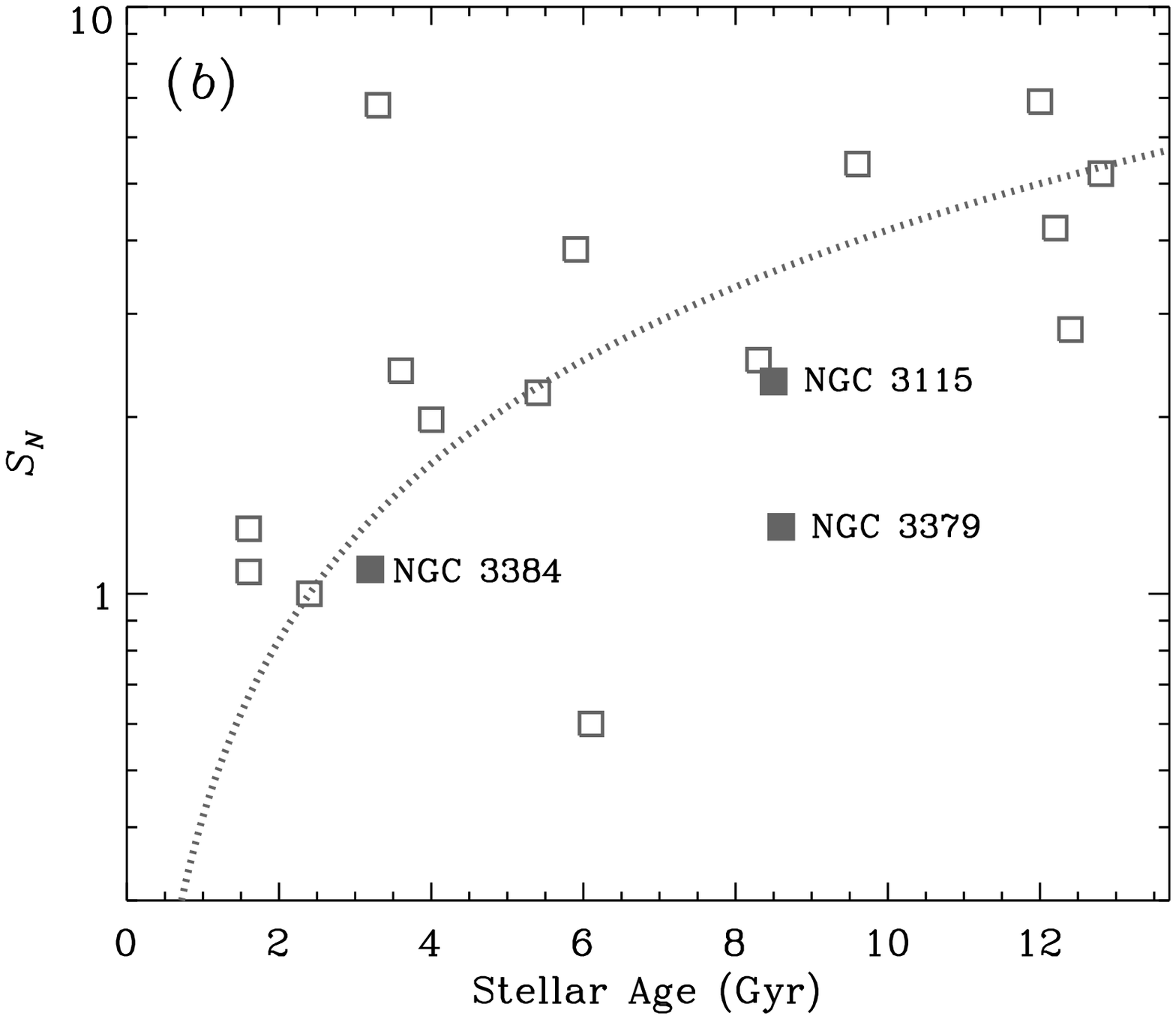}
}
\caption{
($a$) Integrated LMXB \xray\ luminosity per $L_K$ versus stellar age for the
field LMXBs, $(L_{\rm X}/L_K)_{\rm field}$, in the three galaxies in our sample
({\it filled circles with annotations\/}) and the total (GC plus field) LMXBs,
$(L_{\rm X}/L_K)_{\rm total}$, in the Boroson \etal\ (2011) sample ({\it gray
squares\/}.  The three filled gray squares indicate the Boroson \etal\ (2011)
$(L_{\rm X}/L_K)_{\rm total}$ values for the three galaxies in our sample.  The
dashed black curve and surrounding gray region is the expected relation and
spread from the population synthesis models of Fragos \etal\ (2013b), and the
solid curve shows the best-fit version of this curve renormalized to our data.
The dotted gray curve is the best-fit relation to the GC plus field LMXB $(L_{\rm
X}/L_K)_{\rm total}$ versus age data (i.e., the gray squares) and the
long-dashed curve shows how this correlation would change after
subtracting the expected contributions from GC LMXBs (see Eqn.~(7) and
discussion in $\S$~7 for details).
($b$) The observed relation between GC $S_N$ and stellar age with the
best-fitting model ({\it dotted gray curve\/}).
We find that the apparent correlation in Figure~9$a$ between $(L_{\rm
X}/L_K)_{\rm total}$ and stellar age can be fully explained by contributions
from GC LMXBs and the underlying correlation between GC $S_N$ and stellar age
shown in Figure~9$b$.  Furthemore, the contributions from GC LMXBs could mask
an underlying anticorrelation between $(L_{\rm X}/L_K)_{\rm field}$ and stellar
age (e.g., the long-dashed curve in Figure~9$a$).
}
\end{figure*}

The above findings indicate that the young early-type galaxy NGC~3384 contains
a more numerous and \xray\ luminous population of field LMXBs per unit $K$-band
luminosity (stellar mass) than the older early-type galaxies NGC~3115 and 3379
(see $\S$6).  Although the integrated \xray\ luminosity per unit $K$-band
luminosity $L_{\rm X}/L_K$ can be measured to high precision for the LMXB
populations in the three galaxies in our sample, and our measurements clearly
indicate that the field LMXB $(L_{\rm X}/L_K)_{\rm field}$ is larger for
NGC~3384 than for NGC~3115 and 3379, it is important to account for the fact
that the $L_{\rm X}$ value for a given galaxy can be substantially influenced
by the brightest few sources.  If we consider the distribution of point-source
luminosities as a realization of an underlying XLF, then we might expect large
galaxy-to-galaxy variations in $L_{\rm X}/L_K$ simply due to ``statistical
noise'' (see Gilfanov \etal\ 2004 for a detailed discussion on this issue).
For example, a large sample of galaxies with identical underlying XLFs will
contain a spread of $L_{\rm X}$ values due to this statistical noise.

To quantify how this statistical noise influences our results, we made use of a
Monte Carlo technique to measure the expected standard deviation of $L_{\rm
X}/L_K$ for a given XLF.  For each of our LMXB populations in each galaxy, we
constructed 20,000 simulated XLFs that contained distributions consistent with
the best-fitting XLF model (see $\S$5.4 and Table~5).  For a given simulation,
we first computed the expected number of LMXBs that would have $L_{\rm X}
\simgt 10^{36}$~\lum\ and perturbed this value using Poisson statistics.  We
then assigned $L_{\rm X}$ values to each of the sources probabilistically using
the best-fitting XLF model as our underlying probability distribution function.
Finally, the sources within a given simulation were summed to give an
integrated $L_{\rm X}$ value.  The resulting distribution of 20,000 $L_{\rm X}$
values provided us with the effects of the statistical noise on our results.
In Table~6, we have tabulated the integrated $L_{\rm X}$ and $L_{\rm X}/L_K$
values and their errors, which include the statistical noise, and in
Figure~9$a$, we have plotted the field LMXB $(L_{\rm X}/L_K)_{\rm field}$
versus stellar age (and 1$\sigma$ errors) for the galaxies in our sample.  We
have found that the resulting errors on $\log L_{\rm X}$ values are well
described by Gaussian functions, implying asymmetric $L_{\rm X}$ errors.
Considering these errors, and following the error propogation methods described
$\S$2.4.5 of Eadie \etal\ (1971), we find that the field LMXB $(L_{\rm
X}/L_K)_{\rm field}$ for NGC~3384 is a factor of $2.0^{+1.7}_{-0.9}$ and
$3.0^{+2.5}_{-1.4}$ times higher than NGC~3115 and 3379, respectively.  

%
\section{Discussion}
%

Our results support a scenario in which the field LMXB \xray\ luminosity per
unit $K$-band luminosity $(L_{\rm X}/L_K)_{\rm field}$ declines with increasing
stellar age, as predicted by population synthesis models ({\it filled circles}
in Fig.~9$a$).  In Figure~9$a$ we have displayed the expected mass-weighted
stellar age dependent $(L_{\rm X}/L_K)_{\rm field}$ from the population
synthesis models from Fragos \etal\ (2013b) (dashed curve with gray shaded
region).  This curve was derived using equation~(4) of Fragos \etal\ (2013b)
and an assumed $M/L_K \approx 0.66 \; M_{\odot}/L_{K,\odot}$, consistent with
our galaxies.  We find that the population synthesis models predict that
galaxies with mass-weighted stellar ages of $\approx$3.2~Gyr will have $(L_{\rm
X}/L_K)_{\rm field}$ values that are a factor of $\approx$3 times higher than
$\approx$8.5~Gyr galaxies, which is consistent with our finding.  We note that
the absolute scaling of our field LMXB $(L_{\rm X}/L_K)_{\rm field}$ values
appear to be a factor of $\sim$1.5--2 times lower than the maximum likelihood
model adopted in the Fragos models.  However, it is important to note that
these models were calibrated to match local total (i.e., GC plus field)
$(L_{\rm X}/L_K)_{\rm total}$ LMXB scaling relations, which as we have shown in
this study, are affected somewhat by the GC LMXB populations.  Furthermore, the
range of statistically acceptable population synthesis models (i.e., the shaded
region in Fig.~9$a$; see Fragos \etal\ 2013b for details) covers our field LMXB
$(L_{\rm X}/L_K)_{\rm field}$ values.  Using the three $(L_{\rm X}/L_K)_{\rm
field}$ values derived from this study, we adjusted the normalization term in
equation~(4) of Fragos \etal\ (2013b) to fit our results.  Using $\chi^2$
minimization, we found that a factor of $\approx$1.6 downward shift of the
normalization would provide a reasonable fit to our data ({\it solid curve} in
Fig.~9$a$).

Our pilot study provides suggestive evidence that the \xray\ emission from
field LMXB populations declines with stellar age; however, it is absolutely
critical to expand this study to a larger sample of early-type galaxies using
comparable \chandra\ and \hst\ imaging before we draw statistically robust
conclusions.  As we noted in $\S$~1, previous studies have found that the total
(GC plus field) $(L_{\rm X}/L_K)_{\rm total}$ increases with stellar age.  In
Figure~9$a$, we have displayed the $(L_{\rm X}/L_K)_{\rm total}$ versus age for
the Boroson \etal\ (2011) sample.  It is clear that indeed there is a positive
correlation between $(L_{\rm X}/L_K)_{\rm total}$ and stellar age; a Spearman
rank test indicates that the correlation is significant at the $>99.9$\%
confidence level.  However, as pointed out by Zhang \etal\ (2012), this
correlation is strongly influenced by a secondary correlation of GC $S_N$ with
stellar age (see Figure~9b).  As discussed in $\S$6, we have shown that the
$L_K$ and $S_N$ normalized GC XLFs for the galaxies in our sample are
consistent with each other, suggesting that the GC $(L_{\rm X}/L_K/S_N)_{\rm
GC}$ may be a ``universal'' quantity for early-type galaxies generally.  A
similar conclusion was reached by Irwin~(2005), who determined the $B$-band
luminosity normalized quantity ($L_{\rm 0.3-10~keV}$/$L_B$/$S_N$)$_{\rm GC}$~$=
(2.72 \pm 0.47) \times 10^{29}$~\lum~$L_{B,\odot}^{-1}$ for a sample of 12
galaxies.  If we assume $L_{\rm X}/L_{\rm 0.3-10~keV} \approx 0.8$ and $L_B/L_K
\approx 0.18 \; L_{B,\odot}/L_{K,\odot}$, appropriate for our \xray\ binaries
and galaxies, respectively, we can convert the Irwin~(2005) relation to
$(L_{\rm X}/L_K/S_N)_{\rm GC}$~$= (3.70 \pm 0.64) \times
10^{28}$~\lum~$L_{K,\odot}^{-1}$.  Using the data provided in Tables~1 and 6,
we find that $(L_{\rm X}/L_K/S_N)_{\rm GC}$~$\approx 3.0^{+3.5}_{-1.6} \times
10^{28}$~\lum~$L_{K,\odot}^{-1}$ and $1.9^{+4.2}_{-1.3} \times
10^{28}$~\lum~$L_{K,\odot}^{-1}$ for NGC~3115 and 3379, respectively, both
formally consistent with the Irwin~(2005) estimate.  

Given the measurements of GC LMXB $(L_{\rm X}/L_K/S_N)_{\rm GC}$ and
correlations between GC plus field $(L_{\rm X}/L_K)_{\rm total}$ and $S_N$ with
stellar age, we can roughly estimate how the field $(L_{\rm X}/L_K)_{\rm
field}$ would be expected to scale with stellar age.  This can be done by
solving the following system of equations:
\begin{eqnarray}
\log (L_{\rm X}/L_K)_{\rm total} \sim A + B \; t_{\rm age},  \\
(L_{\rm X}/L_K)_{\rm GC} \sim C \; S_N,  \\
S_N \sim D \; t_{\rm age},   \\
(L_{\rm X}/L_K)_{\rm field} = (L_{\rm X}/L_K)_{\rm total} - (L_{\rm
X}/L_K)_{\rm GC}, \nonumber \\
\rightarrow (L_{\rm X}/L_K)_{\rm field} \sim 10^{A + B\; t_{\rm age}} - C
\times D \; t_{\rm age},
\end{eqnarray}
where $A$, $B$, $C$, and $D$ are constants.  Using the Boroson \etal\ (2011)
data presented in Figures~9$a$ and 9$b$, we determined $A \approx
28.84$~$\log$~\lum~$L_{K,\odot}^{-1}$, $B \approx
0.032$~$\log$~\lum~$L_{K,\odot}^{-1}$~Gyr$^{-1}$, and $D \approx
0.42$~Gyr$^{-1}$.  The dotted gray curves in Figures~9$a$ and 9$b$ illustrate
the best-fit solutions to equations~(4) and (6), respectively.  If we adopt $C
\approx 3 \times 10^{28}$~\lum~$L_{K,\odot}^{-1}$ (see above for justification)
and insert this set of values into equation~(7), then we obtain an estimate for
$(L_{\rm X}/L_K)_{\rm field}$, which we show as the long-dashed curve in
Figure~9$a$.  This exercise illustrates that the positive correlation of
$(L_{\rm X}/L_K)_{\rm total}$ with stellar age is dominated by contributions
from GC LMXBs and that it is possible to mask an underlying negative
correlation between $(L_{\rm X}/L_K)_{\rm field}$ and stellar age.  We note,
however, that the above calculation is very sensitive to the set of constants
that are adopted.  Although each of the constants above were derived directly
from the data, it is possible to perturb a constant in a manner entirely
consistent with the data and obtain a qualitatively different result.  For
example, if we adopted $C = 2 \times 10^{28}$~\lum~$L_{K,\odot}^{-1}$, the
value obtained for NGC~3379, we would find the long-dashed curve in Figure~9$a$
to have a flat slope.  Or if we adopted $C = 4 \times
10^{28}$~\lum~$L_{K,\odot}^{-1}$, a value higher than but consistent with those
obtained for NGC~3115 and 3379 and the Irwin~(2005) result, the resulting
$(L_{\rm X}/L_K)_{\rm field}$ from equation~(7) would be negative at $t_{\rm
age} > 6.7$~Gyr.  Therefore, the above calculation should be taken as a proof
of concept and not a well constrained measurement of the stellar age dependence
of $(L_{\rm X}/L_K)_{\rm field}$.  To obtain an accurate measurement of how
$(L_{\rm X}/L_K)_{\rm field}$ varies with stellar age will require direct
measurements of the field LMXB populations in a larger sample of galaxies using
the combination of \hst\ and \chandra\ data sets.  Obtaining such an
accurate measurement will be important for constraining X-ray
binary population synthesis models.  For example, the above inferences for
how $(L_{\rm X}/L_K)_{\rm field}$ changes with stellar age (i.e., the
long-dashed curve in Fig.~9) appears flatter than the current X-ray binary population
synthesis predictions at the youngest stellar ages.  Direct measurements of
$(L_{\rm X}/L_K)_{\rm field}$ for several galaxies are required to show whether
this is indeed the case and whether these predictions will need to be revised.

%
\section{Summary}
%

We have utilized deep \chandra\ and \hst\ observations of the three nearby
early-type galaxies NGC~3115, 3379, and 3384 to identify and study
subpopulations of LMXBs that are found in the galactic field environments.  Our
galaxy sample was selected to span a large range of stellar age
($\approx$3--10~Gyr) so that we could measure how the emission from field LMXBs
depends on stellar population ages.  Our results can be summarized as follows:

\begin{itemize}

\item We find that LMXBs formed dynamically in GCs and unrelated background
\xray\ sources can significantly impact the overall XLF shapes and \xray\ power
output from early-type galaxies, even in low-$S_N$ galaxies like the three
sources studied here ($S_N \simlt 3$).  This finding indicates that for studies
of field LMXBs, it is critical to remove GC LMXBs and unrelated background
sources using the combination of \hst\ and \chandra\ data.  This is especially
the case for the majority of early-type galaxies that have been studied in the
literature, which have $S_N \simgt 2$ (see Fig.~7 and $\S\S$4 and 6).

\item After normalizing for stellar mass (via the $K$-band luminosity), we find
that the field LMXB population in the young ($\approx$3.2~Gyr) galaxy NGC~3384
is more numerous and extends to higher luminosities than the older
($\approx$8.5~Gyr) galaxies NGC~3115 and 3379.  This result is consistent with
previous studies that have found a similar excess of luminous \xray\ sources in
young ($\simlt$6~Gyr) versus old ($\simgt$6~Gyr) early-type galaxies
(e.g., Kim \& Fabbiano 2010; Zhang \etal\ 2012) (see Fig.~7$b$, and $\S$6).

\item Using the field LMXB populations in our galaxies, we measure the
integrated $(L_{\rm X}/L_K)_{\rm field}$ for NGC~3384 to be a factor of
$\approx$2--3 times higher than that found in NGC~3115 and 3379.  This result
is consistent with the trends expected from the \xray\ binary population
synthesis models published by Fragos \etal\ (2013a, 2013b), which predict that
the $(L_{\rm X}/L_K)_{\rm field}$ will be a factor of $\approx$3 times higher
for galaxies with stellar ages of $\approx$3~Gyr compared to those with
$\approx$9~Gyr (see Fig.~9$a$ and $\S\S$6 and 7).

\item We find that the GC LMXB XLFs for NGC~3115 and 3379 are consistent with
each other when normalized by $L_K$ and $S_N$.  We did not detect enough GC
LMXBs in NGC~3384 to study the shape of its GC LMXB XLF; given its low $S_N$,
our lack of GC LMXB detections is not surprising.  We constrained the mean GC
LMXB $(L_{\rm X}/L_K/S_N)_{\rm GC} \approx$~(2--3)~$\times
10^{28}$~\lum~$L_{K,\odot}^{-1}$ for these galaxies, in good agreement with
past studies (Irwin~2005) (see Fig.~8 and $\S\S$6 and 7).

\item Previous studies have found that the total GC plus field LMXB $(L_{\rm
X}/L_K)_{\rm total}$ was positively correlated with stellar age, which was
considered an apparent conflict with population synthesis predictions.  This
observed correlation is largely driven by an underlying correlation between
$S_N$ and stellar age, which masks any underlying correlations of $(L_{\rm
X}/L_K)_{\rm field}$ with stellar age.  We showed that if we consider our
constraint on $(L_{\rm X}/L_K/S_N)_{\rm GC}$ and the $S_N$--stellar age
correlation, it is plausible that $(L_{\rm X}/L_K)_{\rm field} = (L_{\rm
X}/L_K)_{\rm total} - (L_{\rm X}/L_K/S_N)_{\rm GC}$ declines with increasing
stellar age.  However, a robust measurement of $(L_{\rm X}/L_K)_{\rm field}$
versus age will only come from a detailed \chandra\ and \hst\ study of the
field LMXB populations from a larger sample of early-type galaxies (see Fig.~9
and $\S\S$1, 6, and 7).

\end{itemize}

\acknowledgements

We thank the referee for providing thoughtful comments, which have improved the
quality of this paper.  We thank Zhongli Zhang for providing data.  We
gratefully acknowledge financial support from \chandra\ X-ray Center grant
G02-13107A, Space Telescope Science Institute grant GO-12760, and NASA ADP
grant NNX13AI48G (B.D.L.).  F.E.B.  acknowledges support from Basal-CATA
PFB-06/2007, CONICYT-Chile (grants FONDECYT 1101024, Gemini-CONICYT 32120003,
``EMBIGGEN'' Anillo ACT1101), and Project IC120009 ``Millennium Institute of
Astrophysics (MAS)'' funded by the Iniciativa Cientifica Milenio del Ministerio
de Economia, Fomento y Turismo.  W.N.B. acknowledges ADP grant NNX10AC99G.
G.R.S. is supported by an NSERC Discovery Grant. V.K.~acknowledges support for
this work from NASA ADP grant NNX12AL39G (sub-contract to Northwestern U.)

%

%

\end{document}